\documentclass[preprint,11pt,authoryear]{article}

\usepackage{graphicx}
\graphicspath{{.}}

\usepackage{xcolor}
\usepackage[hidelinks,colorlinks=true,linkcolor=blue,citecolor=blue,urlcolor=blue]{hyperref}
\usepackage{fullpage}
\usepackage[right]{lineno}
\usepackage{comment}
\usepackage{subfigure}
\usepackage{amssymb}
\usepackage{mathtools}
\usepackage{authblk}
\usepackage[sort&compress]{natbib}
\usepackage[mediumspace,mediumqspace,squaren]{SIunits}

\usepackage[capitalise,nameinlink,sort&compress]{cleveref}
\crefname{equation}{}{}
\crefname{appendix}{}{}
\creflabelformat{equation}{#2#1#3}
\crefmultiformat{equation}{#2#1#3}{, #2#1#3}{, #2#1#3}{, #2#1#3}
\renewcommand\eqref[1]{(\cref{#1})}

\DeclareMathOperator{\chint}{Chi}
\DeclareMathOperator{\shint}{Shi}

\newcommand{\matr}[1]{\mathbf{#1}}
\renewcommand{\vec}[1]{\boldsymbol{#1}}
\newcommand{\diff}[1]{\mathrm{d}{#1}}
\newcommand{\todo}[1]{{\color{red}{[TODO: \textit{#1}]}}}

\newcommand{\pp}[1]{\left({#1}\right)}
\newcommand{\abs}[1]{\lvert{#1}\rvert}
\newcommand{\bb}[1]{\left[ #1 \right]}
\newcommand{\eqdef}{\vcentcolon =}
\newcommand{\pdiff}[2]
{\frac{\partial{#1}}{\partial{#2}}}
\newcommand{\pdiffn}[3]{\frac{\partial^{#1}{#2}}{\partial{#3}^{#1}}}
\newcommand{\linepdiffn}[3]{{\partial^{#1}{#2}}/{\partial{#3}^{#1}}}

\newcommand{\linepdiff}[2]{{\partial{#1}}/{\partial{#2}}}

\newcommand{\fdiff}[2]{\frac{\diff{#1}}{\diff{#2}}}
\newcommand{\ramp}[1]{\pp{#1}_+}
\newcommand\restr[2]{{
  \left.\kern-\nulldelimiterspace
  #1 
  \vphantom{\big|} 
  \right|_{#2} 
  }}
\newcommand{\E}{\mathrm{e}}
\newcommand{\kon}{k_\mathrm{on}}
\newcommand{\koff}{k_\mathrm{off}}
\newcommand{\ephi}{\vec{{e}_\varphi}}

\newcommand{\halfint}[2]{\int_0^{\infty}{#1 \, \diff{#2}}}

\title{\textbf{The multiscale mechanics of axon durotaxis}}

\author[1]{Christoforos Kassianides}
\author[1]{Alain Goriely\thanks{To whom correspondence should be addressed: \href{mailto:goriely@maths.ox.ac.uk}{goriely@maths.ox.ac.uk}}}
\author[2]{Hadrien Oliveri}

\affil[1]{Mathematical Institute, University of Oxford, Oxford OX2 6GG, United Kingdom}
\affil[2]{Max Planck Institute for Plant Breeding Research, Cologne 50829, Germany}

\begin{document}


\maketitle

 

\begin{abstract} During neurodevelopment, neuronal axons navigate through the extracellular environment, guided by various cues to establish connections with distant target cells. Among other factors, axon trajectories are influenced by heterogeneities in environmental stiffness, a process known as \textit{durotaxis}, the guidance by substrate stiffness gradients.
Here, we develop a three-scale model for axonal durotaxis. At the molecular scale, we characterise the mechanical interaction between the axonal growth cone cytoskeleton, based on molecular-clutch-type interactions dependent on substrate stiffness. At the growth cone scale, we spatially integrate this relationship to obtain a model for the traction generated by the entire growth cone.
Finally, at the cell scale, we model the axon as a morphoelastic filament growing on an adhesive substrate, and subject to durotactic growth cone traction. Firstly, the model predicts that, depending on the local substrate stiffness, axons may exhibit positive or negative durotaxis, and we show that this key property entails the existence of attractive zones of preferential stiffness in the substrate domain. Second, we show that axons will exhibit reflective and refractive behaviour across interface between regions of different stiffness, a basic process which may serve in the deflection of axons. Lastly, we test our model in  a biological scenario wherein durotaxis was previously identified as a possible guidance mechanism \textit{in vivo}. Overall, this work provides a general mechanistic theory for exploring complex effects in axonal mechanotaxis and guidance in general. 
\end{abstract}




\section{Introduction}

The establishment of the neuronal network is a fundamental event in neurodevelopment in which neurons send out slender processes, called \textit{axons}, to meet other target cells--typically other neurons--and physically interconnect with them. During this process, the axons migrates through the extracellular environment towards its functional target, guided by a combination of environmental cues, including gradients in chemical or mechanical properties in their environment. This general process defines \textit{axon guidance}. 

\begin{figure}[ht!]
    \centering
\includegraphics[width=\linewidth]{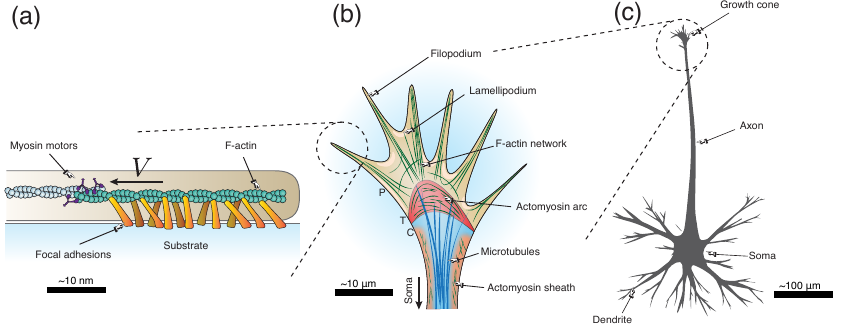}
    \caption{The multiscale structure of the axon. (a) Molecular scale. An actin filament is pulled towards the rear of the growth cone by myosin motors. Actin is coupled mechanically to the substrate via transmembrane adhesive proteins which may detach and attach, producing an overall frictional force. This frictional force is key for axon locomotion. (b) Growth cone scale: schematic representation of the cytoskeleton in a migrating neurite. The shaft (lower part) is essentially composed of cross-linked microtubules surrounded by an actomyosin sheath. The growth cone is a lamellipodium-like structure that confers motility to the neurite and probes the environment. (c) Cellular scale. Schematic of a typical growing neuron showing the soma, the dendrites, and the axon with its growth cone.}
    \label{fig:fig1}
\end{figure}

A developing axon is composed of two main parts (see \cref{fig:fig1}): the \textit{axonal shaft}, a long filament-like section that contains the axoplasm made up of parallel microtubules cross-linked through microtubule-associated tau proteins, neurofilaments, and actomyosin; and the \textit{growth cone}, a lamellipodium-like, actin-supported  structure located at the tip of the axonal shaft \citep{Franze2020,Oliveri2022a}. The growth cone performs both locomotory and sensory functions and, thus, plays a key role in pathfinding. Historically, the focus has been  mostly on \textit{chemotaxis}, the guidance of axons by gradients in the concentration of diffusing chemicals \citep{Mortimer2008}.  The underlying mechanisms and molecular actors of chemotaxis are relatively well described, and have been featured in numerous models \citep{Oliveri2022a}. In contrast, \textit{durotaxis}, the directed motion of cells along gradients in extracellular matrix stiffness \citep{SHELLARD2021227,espina2022durotaxis}, has been identified in axons mostly in the last decade and is only partially understood.

For example, experiments on xenopus optic nerve guidance have suggested the existence of \textit{negative} durotaxis, i.e. migration down the stiffness gradient, from high to low stiffness \citep{koser2016mechanosensing,Thompson2019}. In an apparent contradiction, the same group observed that retinal axons also progress faster on \textit{stiffer} regions \citep[see Fig. 6 in][]{koser2016mechanosensing}. At the scale of an entire axon fascicle, this effect then results  in differential migration speed, hence a deflection of the axonal bundle towards \textit{softer} regions, akin to the bending of light rays in a graded refractive medium \citep{Oliveri2021}. Similarly, axons in chick embryos grow towards softer zones of the somitic segments, strikingly,  even in the context of inhibited  chemotactic signalling \citep{Schaeffer2022,saez2023positive}.  Although mounting evidence supports the presence of neuronal durotaxis in vivo, its specific role and underlying mechanisms remain unclear.

By definition, durotaxis is a mechanics-based mechanism. Hence, a complete description of durotaxis must rely on the mechanical effects underlying growth and locomotion. The precise nature and mechanism of axonal growth have been long debated and remain controversial  \citep{Franze2020,Oliveri2022a,Suter2011}. Generically, \textit{growth} refers to the general process by which a body changes form  by virtue of a change in mass \citep{Goriely2017}. In axons, the emerging view is that traction forces generated by the growth cone create mechanical tension along the trailing shaft; the latter then elongates as an effect of these forces. Thus, we define \textit{axonal growth} as the {irreversible} elongation of the axonal shaft resulting from these traction forces and supported by material addition \citep{Suter2011,OToole2008,nguyen2013tension,Holland2015,Recho2016,purohit2016model,miller2018integrated,Franze2020,Oliveri2022a,Oliveri2022b,Goriely2015}. Microscopically, the mechanical forces generated by the growth cone arise from the mechanical coupling between the growth cone's actomyosin network and the substrate \citep{Franze2020,RAFFA2022,Chan2008,isomursu2022directed}. The effect of matrix compliance in this process can be understood through the so-called \textit{molecular-clutch hypothesis} \citep{elosegui2018control}. Within the growth cone, actin undergoes a sustained retrograde flow, from the periphery of the growth cone to the base, generated by active myosin motors located at the base of the growth cone. This actin flow, in turn, generates micro-frictional interactions between the actin and the substrate, mediated by transmembrane adhesion complexes that may form and detach stochastically. 

Overall, our general problem is to establish a physical law of motion for axons embedded in a heterogeneous mechanical environment. Here, we focus on two main problems. The first  problem is  to understand how these molecular interactions between substrate and axon result in the generation of an overall force. Then, from the molecular mechanics of growth cone locomotion, the second problem is to understand how this force affects the overall trajectory of the axon.  

\begin{figure}[ht!]
    \centering
\includegraphics[width=\linewidth]{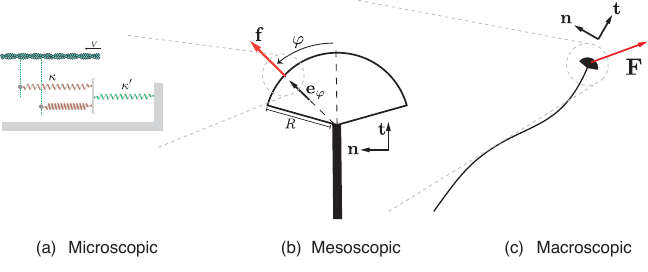}
    \caption{General structure of the model. From left to right: microscopic scale--an actin filament is modelled as a rigid rod moving with velocity $V$ w.r.t. the substrate. The filament is connected to the substrate through an evolving collection of springs modelling adhesive cross-bridges and the deformable substrate; mesoscopic scale--the growth cone is viewed as a continuous distribution of centripetally-aligned actin filaments with individual traction forces summing up to generate global net force on the axonal shaft; macroscopic scale--the axon is modelled as a morphoelastic filament subject to frictional forces between the shaft and the substrate and growth-cone-induced locomotory force.}
    \label{fig:model}
\end{figure}

To address these problems,  we develop a multiscale, mechanical model for axonal durotaxis, summarised in \cref{fig:model}. 
We start with the molecular scale by describing the mechanical balance of a single actin filament interacting with a substrate (\cref{micro}) and we derive the one-dimensional locomotory force resulting from this interaction. At the mesoscale (\cref{meso}), we integrate the contributions of all actin filaments forming the growth cone actomyosin meshwork to compute the net force produced by the growth cone. Finally, at the macroscale (\cref{macro}), this growth cone  locomotory force is integrated in a rod theory to predict the trajectory of the whole axon.

\section{The microscopic model\label{micro}}

\subsection{Adhesion of one actin filament}

 The mechanical interaction between the substrate and the cytoskeleton is mediated by adhesion proteins, henceforth generically termed \textit{cross-bridges}. In the growth cone, tension is generated by the movement of actin which stretches these cross-bridges. This tension opposes the retrograde motion of actin and pulls on the substrate, resulting reciprocally in a reaction force acting upon the axon shaft. The cross-bridges randomly bind and unbind, with a rate that depends on their level of tension \citep{bell1984cell,Chan2008}. Overall, random detachment and attachment of cross-bridges, combined with stretch-induced tension result in emergent frictional interactions between the actin and the substrate which  depends on substrate stiffness \citep{Chan2008,sens2013rigidity}. The emergence of this frictional force is the central mechanism considered here underlying growth cone locomotion.

\begin{figure*}[ht!]
     \centering
\includegraphics[width=0.5\linewidth]{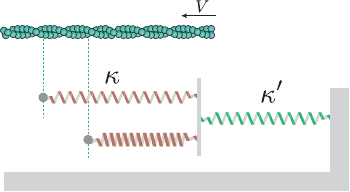}
    \caption{Structure of the microscopic model. We assume that all cross-bridges  (brown springs with constant $\kappa$) interact with the same substrate spring (green spring with constant $\kappa'$).}
    \label{fig:protein schematic}
\end{figure*}

To model the mechanical interaction between a single actin filament and the substrate, we adapt the approach of \cite{sens2013rigidity}. The filament is modelled as a straight, infinite rigid rod moving relative to the substrate with velocity $V$, as shown in \cref{fig:protein schematic} (we also introduce the longitudinal position $X=Vt$). In the steady regime, $V$ is constant in time. We model the cross-bridges as linearly elastic springs connecting the actin filament to the substrate. 
Each spring has an individual tension $f$, elongation $x$, and spring constant $\kappa$ related through Hooke's law by
\begin{equation}
    f = \kappa x.
\end{equation}
The cross-bridge tension can be related to the time $\Delta t$ elapsed since its attachment. Indeed, for a perfectly rigid substrate, 
the tension of a single cross-bridge since attachment is $f \approx \kappa V \Delta t$ (assuming that cross-bridges form unstretched, and neglecting the geometric nonlinearity due to the distance between the actin filament and the substrate). The situation is more complex in the case of a deformable substrate, where the cross-bridges may interact indirectly and non-locally with one another through short- and long-range deformations of the substrate \citep{nicolas2008dynamics}. This deformation may be expressed in terms of the Green solution for an infinite half-plane subject to surface traction \citep{Hwu2010,sens2013rigidity,nicolas2008dynamics}.

Here, for simplicity, we assume that the substrate is soft so that  cross-bridges interact mechanically as the substrate deforms almost uniformly. The substrate is then modelled as a single spring $\kappa'$ with which all cross-bridges   are connected in parallel. To model the kinetics of cross-bridge detachment, we posit a Bell-type law \citep{Bell1978}, i.e. we assume that the probability that a cross-bridge detaches in a small time-interval  $\Delta t$ is related to its tension $f$ according to
\begin{equation}
    p_{\mathrm{off}} = \koff \Delta t \, \E^{f / f_0},
\end{equation}
where $f_0$ is the characteristic force scale of detachment and $\koff$ is the kinetic off-rate at rest. Further, we assume that cross-bridges form in the unstretched state at a constant rate $\kon$, so the probability of attachment during $\Delta t$ is
\begin{equation}
    p_{\mathrm{on}} = \kon \Delta t.
\end{equation}
Finally, we focus on a small section of the filament at the growth cone edge. Therefore, we assume that  all material parameters are constant in space which reduces the  problem to a zero-dimensional problem.

\subsubsection{Mean-field treatment}
The dynamics of focal adhesions has been previously analysed using continuum or computational approaches \citep[e.g.][]{srinivasan2009binding, sabass2010modeling, bressloff2020stochastic}, in particular with a focus on the role of substrate compliance in \citep{bangasser2013master,Chan2008,sens2013rigidity}. 
Similar to the mean-field approaches of \cite{bangasser2013master,sens2013rigidity} we here adopt a continuum perspective where we view the set of bound cross-bridges and their respective mechanical state as a continuum quantity. That is, we first consider the distribution of the extensions of bound cross-bridges. Let $n$ be the density of bound cross-bridges per extension level, such that, at time $t$ there are (on average) $\diff N = n\pp{x, t} \diff x$
bound cross-bridges with extension in  $\bb{x, x+\diff x}$. This density obeys the Lacker-Peskin equation \citep{srinivasan2009binding}
\begin{equation}\label{eqn:Lacker-Peskin}
    \pdiff{n}{t} + v(x,t) \pdiff{n}{x} = B(x,t) - U(x,t),
\end{equation}
where $B(x,t)$ and $U(x,t)$ denote, respectively, the kinetic binding and unbinding rates for cross-bridges at extension $x$; and $v\pp{x,t}$ is the extension rate of bound cross-bridges. Assuming a large, constant pool of free (unbound) cross-bridges with concentration $c_\mathrm{free}$, binding occurs at constant rate $S_0 = \kon c_\mathrm{free}$ (with unit mole per unit time). Assuming that cross-bridges attach with zero extension, we posit an attachment rate of the form \begin{equation}\label{eqn:Bdef}
    B(x,t) = S_0 \delta(x) ,
 \end{equation}
with $\delta$ denoting the Dirac delta function. The Bell-type law for cross-bridge detachment can be expressed as \citep{sens2013rigidity}
\begin{equation}\label{eqn:Udef}
    U(x,t) = \koff \, \E^{\kappa x / f_0} n(x,t).
\end{equation}
The extensions $ x$ and $x_{\mathrm{s}}$ of a cross-bridge and the substrate, respectively, obey the virtual kinematic relation
\begin{equation}\label{eqn:kinematics}
     \delta {x} + \delta {x}_{\mathrm{s}} - \delta X = 0, 
\end{equation}
where, here $\delta {x}$, $\delta {x}_{\mathrm{s}}$ and $\delta {X}$ denote virtual displacements of the cross-bridge, the substrate and the actin filament, respectively.

 Since we assumed previously that all cross-bridges are connected to the same substrate spring ($\kappa'$), the force balance is 
\begin{equation}
    \kappa' x_\mathrm{s} = \int_0^\infty \kappa x n\,\diff x.
\end{equation}
In a virtual displacement of the filament we have
\begin{equation}
    \kappa' \delta x_\mathrm{s} = \int_0^\infty{\kappa \delta x n}\,\diff x.
\end{equation}
Using \eqref{eqn:kinematics} to remove $\delta x_\mathrm{s}$ we obtain
\begin{equation}
    \chi (V\delta t -  \delta x) = N \delta x,
\end{equation}
where  $\chi \eqdef \kappa' / \kappa$ is a dimensionless measure of the substrate stiffness; and where the total amount of bound cross-bridges is 
\begin{equation}\label{N}
    N  = \halfint{n}{x}.
\end{equation}
Since $v\delta t= \delta x$, we obtain
\begin{equation}
  v =  \frac{V}{1+N/\chi},\label{vv}
\end{equation}
and we can rewrite the governing equation for $n$, \eqref{eqn:Lacker-Peskin}, as
\begin{equation}
  \pdiff{n}{t} +  \frac{V}{1+N/\chi} \pdiff{n}{x} =  S_0 \delta(x)- \koff \, \E^{\kappa x / f_0}n(x,t) .
  \label{eqn: n equation correlated}
\end{equation}

\subsubsection{Steady-state traction}

Henceforth, we restrict our attention to time-independent solutions of the problem so we set $\linepdiff{n}{t}=0$ in \eqref{eqn: n equation correlated}. 
The total force $T$ exerted by the cross-bridges on the actin filament is  
\begin{equation}
    T  =   \halfint{n  \kappa x}{x},
\end{equation}
which, after integration of \eqref{eqn: n equation correlated} for $n(x)$, leads to
\begin{equation}
    T\pp{\chi,\nu} =  f_0 K  \,\mathcal{H}\pp{\frac{q}{\chi \nu}\pp{\chi + N}}.\label{eqn:Texpression}
\end{equation}
Here the dimensionless parameters $K \eqdef S_0 / \koff$, $q \eqdef f_0 / \kappa \ell$ and $\nu \eqdef V / (\koff \ell)$ denote respectively the binding constant of cross-bridges, their bond strength, and a dimensionless measure of actin velocity; and $\mathcal H$ is the special function defined as
\begin{equation}
  \quad \mathcal H \pp{x} \eqdef 
  x \E^x \int_0^\infty y \exp\pp{- x \E^y}\,\diff y, \quad x> 0.
\end{equation}
The total number of cross-bridges, $N$, defined in \eqref{N} and present in \eqref{eqn:Texpression}, satisfies  at  steady state the equation
\begin{equation}
    N = \frac{K}{ \nu}\halfint{\exp\left[\frac{q(\chi + N)}{\chi \nu}\pp{1 - \exp\left(\frac{\chi x}{q(\chi + N)}\right)}\right]}{x}.
    \label{eqn: num proteins}
\end{equation}

\subsubsection{Force balance}

We have derived in \eqref{eqn:Texpression,eqn: num proteins} a relation linking traction generation with actin velocity and substrate stiffness. However, the motion of actin itself results from the activity of myosin motors at the rear of the growth cone, thus, the dimensionless velocity $\nu$ is not prescribed, but is instead  an unknown that is obtained from the force balance between actin and the myosin motors. 

The forces acting on a single actin filament are generated by myosin contraction \citep{lin1996myosin} modelled as a constant force $M$. In addition, actin is subject to viscous friction with the cytoplasm (with constant $\Xi$), as well as the traction force generated by the cross-bridges, $T\pp{\chi, \nu}$ given by \eqref{eqn:Texpression}. Thus, neglecting inertial effects, the force balance reads
\begin{equation}
    M - T\pp{\chi, \nu} - \Xi \, V  = 0. \label{eqn:force balance dimensional}
\end{equation}
We define
\begin{equation}
     u \eqdef  \frac{q  }{ \nu}\pp{ 1 + \chi^{-1}N  },
     \quad \beta\eqdef\frac{ \koff \, \Xi \, f_0}{\kappa},
\end{equation}
to remove the dependence upon $\nu$ from \eqref{eqn:force balance dimensional,eqn: num proteins}:
\begin{subequations}
\label{eqn:force-balance-u-Nu}
\begin{align}
    \chi \, u \pp{M - f_0 K \mathcal{H}\pp{u}} - \beta \pp{ \chi +  N } = 0, \label{eqn:force-balance-u}
   \end{align}
   \begin{align}
   N = \frac{K u \chi}{ q \pp{\chi + N}}\halfint{\exp\left[u\pp{1 - \exp\left(\frac{\chi x}{q(\chi + N)}\right) }\right]}{x},\label{eqn:Nu}
\end{align}
\end{subequations}
where we have used \eqref{vv}.
 For a given $\chi$, \eqref{eqn:force-balance-u-Nu} forms a system of equations for the unknowns $\pp{u, N}$. The solution to that system can be substituted into \eqref{eqn:Texpression} to obtain the traction
\begin{equation}
    T\pp{\chi } = f_0 K \mathcal{H}\pp{u},
    \label{eqn: traction u}
\end{equation}
where $u$ is viewed as an implicit function of $\chi$. This expression relates the traction force to the substrate stiffness directly, in contrast to the work of \cite{sens2013rigidity} where the traction force is obtained for a given actin velocity $\nu$. 
\subsection{Optimal stiffness}
As can be seen, for a range of parameters, the traction-stiffness function $T\pp{\chi }$  may exhibit global maximum at a stiffness $\chi^*$. We call $\chi^*$ the \textit{optimal stiffness} defined as the stiffness where traction is maximal $T_\text{max}=T\pp{\chi }$. At that value, the ability of the axon to develop traction forces on the substrate is maximal. From \eqref{eqn: traction u} we see that $T\pp{\chi}$ is maximal when $\mathcal{H}\pp{u\pp{\chi}}$ is  maximal. The function $\mathcal{H}\pp{x}$ has a global maximum at $x = x^* \approx 0.387$ (at which point $\mathcal H (x^* )\approx 0.298$), and therefore $T\pp{\chi}$ is maximised when $u\pp{\chi} = x^*$. 

Substituting $u = x^*$ and $N=N^*$ into \eqref{eqn:force-balance-u-Nu} and rearranging the terms, we obtain
\begin{subequations}
    \begin{equation}
   \chi^* = \frac{\beta N^*}{x^*\pp{M - f_0 K \mathcal{H}\pp{x^*}} - \beta},
    \label{eqn: chi star}
    \end{equation}
    \begin{equation}
        N^* =  \frac{K x^* \chi^*}{ q \pp{ \chi^* +  N^*}} \halfint{\exp\left[x^*\pp{1 - \exp\left(\frac{\chi^* \, \xi}{q(\chi^* + N^*)}\right)}\right]}{\xi}.
    \label{eqn: N star}
    \end{equation}
\end{subequations}
This defines a system of two equations for the two unknowns $\pp{\chi^*, N^*}$. Remarkably, there exists a closed-form solution for this system given by
\begin{equation}\label{eqn:chistar-gamma1}
    \chi^* = \frac{\beta K x^* \E^{x^*} \Gamma\pp{0, x^*}}{x^*\pp{M - f_0 K \mathcal{H}\pp{x^*}} - \beta},\quad
    N^* 
=    K x^* \E^{x^*}\Gamma\pp{0, x^*},
\end{equation}
with $\Gamma\pp{s,x}$ the upper incomplete gamma function.

Since $N^*$  is positive if $\chi^*$ is  positive, we see from \eqref{eqn:chistar-gamma1} that the condition for the existence of a positive local maximum $\chi^*$ is
\begin{equation}
    x^*\pp{M - f_0 K \mathcal{H}\pp{x^*}} - \beta > 0.
    \label{inequality: durotaxis condition}
\end{equation}
 When this condition is not satisfied, the traction is a strictly increasing function of $\chi$ with an horizontal asymptote  as $\chi \rightarrow \infty$.
Importantly, we will show in \cref{meso} that the emergence of negative durotaxis is directly predicated upon the existence of a positive optimal stiffness via the condition \eqref{inequality: durotaxis condition}. 
Moreover, the second derivative $T''(\chi^*)$ of the traction $T(\chi)$ at the optimal stiffness $\chi^*$ characterises the behaviour of the system near this optimum. 
Differentiating \eqref{eqn: traction u} twice with respect to $\chi$ and setting $\chi = \chi^*$ we obtain 
\begin{equation}
    T''\pp{\chi^*} = f_0 K \mathcal{H}''\pp{x^*} {u'(\chi^*)}^2.
    \label{eqn:curvature T}
\end{equation}
An exact expression for $  u' \pp{\chi^*}$ in the previous equation is obtained by differentiating \eqref{eqn:force-balance-u-Nu} w.r.t. $\chi$ at $\chi = \chi^*$, providing a system of two equations for the unknowns $  u '\pp{\chi^*}$ and $   N' \pp{\chi^*}$, which can be solved to obtain
\begin{equation}
  u'(\chi^*) = \bb{1 + \frac{1}{x^*}\pp{\frac{\E^{x^*}}{ {\chint\pp{x^*} - \shint\pp{x^*}}}  + \frac{  \koff \Xi}{  { \koff \Xi +    x^* \pp{K \kappa\mathcal{H}\pp{x^*}- M  \kappa/f_0}}}}}^{-1},
    \label{eqn:derivative u}
\end{equation}
with $\chint$ and $\shint$ the hyperbolic cosine and sine integrals, respectively. 

\subsection{The effect of myosin contractility\label{subsection:contractility}}

The parameter $M$  quantifies the force of myosin contraction, which plays an important role in axon guidance \citep{mccormick2020mechanistic, turney2005laminin}. 
The parameter $M$ changes qualitatively  the traction-stiffness curve. Indeed, we define $M_\text{crit}$ the value of $M$ at which the inequality \eqref{inequality: durotaxis condition} is an equality, that is,
 \begin{equation}
    M_\text{crit}= f_0 K \mathcal{H}\pp{x^*} +\beta/x^*. 
 \end{equation}
This critical force defines two distinct behaviours. 

\begin{figure}[ht!]
    \centering
    \includegraphics[width=0.8\linewidth]{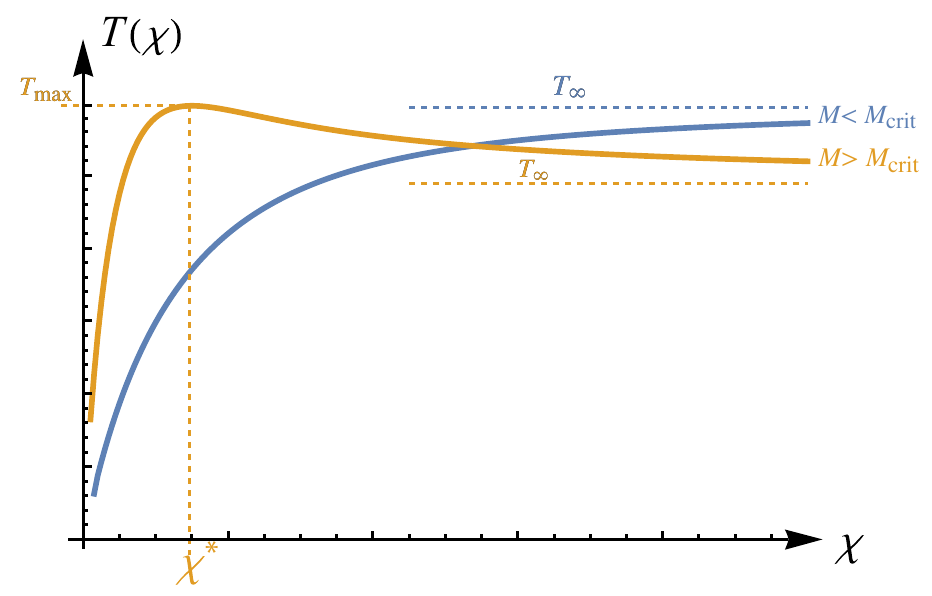}
    \caption{Profile of the traction-stiffness curve $T\pp{\chi}$. For small myosin concentration, $M<M_\text{crit}$, the traction increases monotonically as a function of the stiffness. For large myosin concentration, $M>M_\text{crit}$, traction has a maximum value at the optimal stiffness $\chi^*$.}
    \label{fig:protein traction general}
\end{figure}

First, for $M>M_\text{crit}$,  when myosin contraction is sufficiently large, as shown in \cref{fig:protein traction general}, there exists an optimal stiffness and  the behaviour of the system is controlled by three important features of the traction-stiffness curve:
(i) the optimal  stiffness $\chi^*$;
(ii) the \textit{curvature} $T'' \pp{\chi^*}$ defining how quickly traction changes close to the optimal stiffness; (iii)
the \textit{asymptotic traction} generated on a rigid substrate $T_\infty \eqdef \lim_{\chi\rightarrow \infty} T\pp{\chi}$.
 We note that the optimal stiffness $\chi^*$ is a decreasing function of $M$ and tends to zero as $M \rightarrow \infty$; see \eqref{eqn:chistar-gamma1}. As can be seen from \eqref{eqn:curvature T,eqn:derivative u}, $T''\pp{\chi^*}$ is a bounded, increasing function of $M$ such that
\begin{equation}
   T''\pp{\chi^*} \mathop{\rightarrow}_{M\rightarrow \infty} f_0 K \mathcal{H}''\pp{x^*} \pp{{1-\frac{\E^{-x^*}}{x^*\Gamma\pp{0, x^*}}  }}^{-2}
\end{equation}
 This asymptotic value is  the maximal curvature achievable at $\chi^*$ by modulating contractility $M$. We see that the curvature increases as $\chi^*\to 0$ as $M\to\infty$. However, there is a physical bound to the value of $M$ depending on available binding sites. This relationship shows that the larger the myosin contraction, the faster an axon will react to changes in stiffness close to the optimal stiffness.
  
 To investigate the behaviour of $T_\infty$ in the limit $M \rightarrow \infty$, we first divide \eqref{eqn:force-balance-u} by $\chi$ and take the limit $\chi \rightarrow \infty$ to obtain
\begin{equation}
    u\pp{M - f_0 K \mathcal{H}\pp{u}} - \beta = 0.
    \label{eqn: u equation 2}
\end{equation}
Note that $T_\infty = f_0 K \mathcal{H}\pp{u}$ in the above equation and therefore we are concerned with the asymptotics of $\mathcal{H}\pp{u}$ in the limit $M \rightarrow \infty$. 
Showing that $ u \sim \beta /M $ as $M\rightarrow\infty$,
it follows that 
   $ T_\infty   
   \sim f_0 K \mathcal{H} \pp{  {\beta}/{M}} \rightarrow 0 $.

Second, when myosin contraction is weak, that is for $M<M_\text{crit}$, there is no maximum and traction is a monotonically increasing function of $\chi$, plateauing to $T_\infty = \lim_{\chi\rightarrow \infty} T\pp{\chi}$. Noting that $u \sim \pp{\beta + f_0 K}/M$ as $M \rightarrow 0$ we obtain 
\begin{equation}
    T_\infty \mathop{\sim}_{M \rightarrow 0} f_0 K \mathcal{H}\pp{\pp{\beta + f_0 K}/M} \sim f_0 K M/\pp{\beta + f_0 K}. 
\end{equation}
We see that as $M$ increase past $M_\text{crit}$, $T_\infty$ decreases. Hence, for large myosin contraction, axons deposited on a very rigid substrate will  exert a reduced traction. 

\section{The mesoscopic model\label{meso}}
Macroscopically, the total force generated by the growth cone results from the centripetal flow of all actin filaments organised radially in the lamellipodium towards the growth cone base. We assume that the specific shape of the growth cone has little effect on the net force generated by the cone. Hence  we model it  as a circular sector with central angle $2\varphi_0$ and radius $R$; see \cref{fig:model}.  Here, we denote by $\vec r$ the position  of the base of the growth cone in the laboratory frame equipped  with the canonical basis $\left\{\vec e_x, \vec e_y\right\}$. We denote by  $\vec t$ the unit vector along the axis of symmetry of the growth cone that is also the tangent to the axon attached to the growth cone, and $\vec n$ the normal unit vector to $\vec t$. We define $\vec e_\varphi = \cos \varphi \,\vec t + \sin\varphi \,\vec n,$ with $ \varphi\in[-\varphi_0,\varphi_0]$, the unit vector pointing outward in the direction $\varphi$.

The force density applied at the edge of the disk along the direction $\varphi$ depends on the local substrate stiffness at the lamellipodium edge at this location, i.e. at the point $\vec{r} + R\ephi$.
Neglecting the mutual mechanical interactions between the filaments, the resultant force applied at the distal end of the axon shaft is given by
\begin{equation}
      \vec F\pp{\vec r,\theta} = \int_{-\varphi_0}^{\varphi_0} \rho T\pp{\chi\pp{\vec{r} + R\ephi}} \ephi\, R\,\diff{\varphi}.\label{eqn:gc_resultant_integral}
\end{equation}
with $\rho$ the density of adhesion sites per unit arclength along the growth cone edge (which we assume to be constant); and
where $T(\chi)$ is the individual filament force given by \eqref{eqn: traction u} evaluated at the edge of the growth cone where adhesion sites are located. 
Note that, since all the filaments point towards the base of the growth cone, there is no induced torque about that point, thus changes in direction may only be induced through an asymmetric distribution of forces across the growth cone. The expression \eqref{eqn:gc_resultant_integral} is difficult to work with as it implies evaluating $T$ for every $\varphi$. To make progress, we assume that the stiffness field varies slowly in the vicinity of the growth cone--equivalently that the growth cone is small--so we expand \eqref{eqn:gc_resultant_integral} to $O(R^2)$ to obtain
\begin{equation}
    \vec F 
        =
         2 R\rho \sin{\varphi_0} T\pp{\chi} \vec t + R^2 \rho\varphi_0
      T'\pp{\chi}\matr{P}\pp{\varphi_0}\nabla\chi +  O (R^3) .\label{eqn:force-gc-linearised}
\end{equation}
where $\chi$ and $\nabla \chi$ are evaluated at  $\vec r$; and where the tensor $\matr P$ reflects the geometry of the growth cone only and is given by 
\begin{align}\label{eqn:definitionP0}
    \matr{ P}\pp{\varphi_0} \eqdef    \int_{-\varphi_0}^{\varphi_0}\ephi\otimes\ephi\,\frac{\diff{\varphi}}{\varphi_0}  = \matr 1 + \frac{\sin 2\varphi_0}{2\varphi_0}\pp{\vec t\otimes \vec t - \vec n\otimes\vec n},
\end{align} 
Here, $\matr 1$ is the identity tensor, and `$\otimes$' denotes the tensor product (i.e. given vectors $\vec u$ and $\vec v$, the Cartesian components of $\vec u\otimes \vec v$ are  $(\vec u \otimes \vec v)_{ij}=u_iv_j$). From this expansion, the role of substrate stiffness and its gradient can be interpreted physically. 

The first term on the r.h.s. of \eqref{eqn:force-gc-linearised} is a stiffness-dependent force component in the current direction of the axon, regulating the longitudinal growth velocity. The second term captures the durotactic effect induced by a gradient in substrate stiffness $\nabla \chi$, which will cause the axon to steer away or towards stiffness gradients. In the special cases $\varphi_0=\pi/2$ (half-disk) or $\varphi_0=\pi$ (full disk), the durotactic component is aligned with $\nabla \chi$ as $\matr{P}\pp{\pi}=\matr{P}\pp{\pi/2}=\matr 1$. More in general, the durotactic force component is not aligned with $\nabla\chi$ as $\nabla\chi\times\matr P\nabla\chi\neq \vec 0$. However, since $\matr P$ is symmetric positive, we have that  $\nabla\chi\cdot\pp{\matr{P}\nabla\chi}\geq 0$, thus, if $T'(\chi)>0$, the durotactic force makes an acute angle with the stiffness gradient. Consequently, the orientation of the durotactic force, that is, whether the axon progresses towards soft or stiff substrate, is only governed by the sign of $T'\pp{\chi }$. We have \textit{positive durotaxis} if $T'\pp{\chi }>0$ and \textit{negative durotaxis} if $T'\pp{\chi }<0$. Note that, as expected, the normal component of the durotactic force vanishes when the axon is aligned with the gradient. 

From \eqref{eqn:force-gc-linearised}, we can directly obtain the factors determining the strength of the durotactic force. We define the dimensionless \textit{durotactic sensitivity} as
\begin{equation}
   \vartheta(\chi)=  \frac{\varphi_0}{\sin\varphi_0} \frac{ T'\pp{\chi}}{ T(\chi)},
\end{equation} 
measuring the relative effect of the deflective force to the longitudinal traction.  From  \cref{micro}, it follows that  $\vartheta$ vanishes at the optimal stiffness $\chi^*$, the stationary point of $T$. For $\chi < \chi^*$, we have  $\vartheta>0$ for $\chi < \chi^*$ and, conversely,  $\vartheta<0$ for $\chi > \chi^*$. 

From the mesoscopic model,  we conclude that: \textit{Axons growing on substrates with stiffness smaller than the optimal stiffness $\chi^*$ can only exhibit positive durotaxis; conversely,  axons on substrates stiffer than the optimal stiffness can only exhibit negative durotaxis}. In particular, if condition \eqref{inequality: durotaxis condition} is violated, axons can only undergo positive durotaxis.

\section{The macroscopic model}\label{macro}

In the previous two sections, we have derived the overall net force applied to the tip of the growing axon due to the interaction of the growth cone with the substrate. We can now use this force and model the axon as a morphoelastic rod, i.e. a growing elastic rod \citep[see Chap. 5 and 6 of][]{Goriely2017} subject to the end loading due to the growth cone and to the frictional force exerted by the substrate, thus generalising the approach of \cite{OToole2008} to curved trajectories. 

\subsection{Morphoelastic rod model\label{morphorod}}
Initially, a morphoelastic rod is defined by a smooth curve  $\vec R_0\pp{S_0} = \pp{X_0\pp{S_0}, Y_0\pp{S_0}} \in\mathbb{R}^2$ with arclength $S_0\in \bb{0, L_0}$. We choose $S_0=0$ to be at the base (the soma)  and $S_0=L_0$ at the axon tip (growth cone).
As the axon grows and deforms elastically under loads, the rod adopts a new \textit{current} configuration parameterised by the material coordinate $S_0$ and time $t$ by $\vec r \pp{S_0, t} = \pp{x \pp{S_0, t},y \pp{S_0, t}}$. We define $s\pp{S_0,t}$ the arclength in the current configuration, i.e. the contour length between the base and the material point that was at position $S_0$ and time $t=0$. The Frenet-Serret frame of the current curve is composed of the unit tangent vector $\boldsymbol{\tau}=\linepdiff{\vec r}{s}$ (pointing towards the growth cone) and the unit normal vector $\boldsymbol{\nu}$, perpendicular to $\boldsymbol \tau$. Both can be encoded by a single angle $\theta$ between the $x$-axis and the tangent vector so that  
\begin{align}
\boldsymbol{\tau}=\pp{\cos\theta,\sin\theta},\quad \boldsymbol{\nu}=\pp{-\sin\theta,\cos\theta}.
\end{align} 
At the tip, the orientation of the axon agrees with that of the growth cone so that $\boldsymbol{\tau}(L_0)=\vec t$ and $\boldsymbol{\nu}(L_0)=\vec n$.
Then the Frenet-Serret equations can be written as
\begin{equation}
\pdiff{x}{S_0} = \lambda \cos\theta, \quad \pdiff{y}{S_0} = \lambda \sin\theta, \quad \pdiff{\theta}{S_0}= \lambda \kappa, \quad \lambda= \pdiff{s}{S_0},\label{eqn:kinematic_relations}
\end{equation}
where $\lambda$ is the  stretch and $\kappa$ the signed curvature. The pair of scalar fields $\pp{\lambda, \kappa}$ then fully describes the geometry of the rod (stretching and bending).
\begin{figure*}[ht!]
    \centering
\includegraphics[width=.8\linewidth]{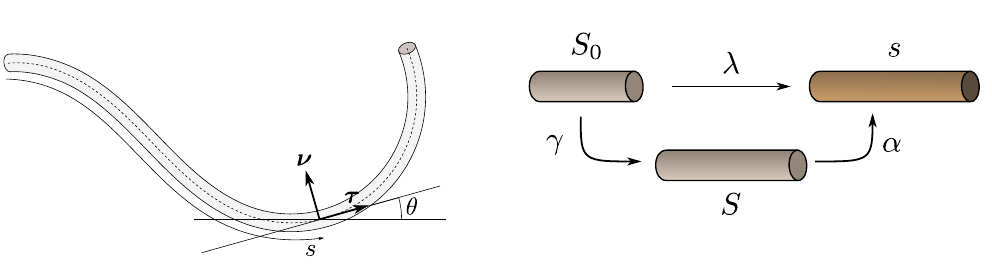}    
    \caption{A morphoelastic rod model. Left: At a point $s$ in the current configuration, we defined the tangent vector $\boldsymbol{\tau}$ and normal vector $\boldsymbol{\nu}$ through the angle $\theta$. Right: In the growth model we assume that a infinitesimal element of the rod located initially at $S_0$ is subject to both a growth deformation $\gamma$ and an elastic deformation $\alpha$, such that the total stretch is $\lambda=\alpha\gamma$.}
    \label{fig:rod}
\end{figure*}

Neglecting inertial effects and  body torques, the balance of linear and angular momenta is expressed in terms of the internal resultant force $\vec n=(n_x,n_y)$ and moment $m $ applied by a cross-section $S_0^+$ onto a cross-section $S_0^-$ as \citep[see chapter 5 of][for details]{Goriely2017}
\begin{subequations}\label{eqn:balance}
    \begin{align}
&  n_x' + \lambda   b_x = 0,\label{eqn:nbalancex}\\
&  n_y' + \lambda b_y = 0,\label{eqn:nbalancey}\\
& m' +  n_y\cos \theta  - n_x \sin\theta    =   0 , 
\label{eqn:mbalance}
\end{align}
\end{subequations}
where the apostrophe denoting differentiation w.r.t. $S_0$; and
 $\vec b=(b_x,b_y)$ is a density of  body force per unit current length, that is used to model the adhesion between the growing shaft and the substrate  as a linear frictional force with coefficient $\zeta$ \citep{OToole2008,Oliveri2021,Oliveri2022a}
\begin{equation}
    \vec b = - \zeta  \vec {\dot r} .\label{eqn:adhesionforce}
\end{equation}
Here the overdot denotes the material derivative, linked to the Eulerian derivative through the formula $\vec {\dot r} - \dot s \boldsymbol \tau= \linepdiff{\vec r(s,t)}{t}$.

To close the system, we must formulate constitutive relationships that relate stresses to deformations. We assume that the rod is  extensible but unshearable. In a growing rod, only part of the deformation results in stresses, reflecting the remodelling and growth of the material and the resulting change in reference configuration. To model this effect, we introduce an intermediate reference configuration $\vec R \pp{S_0,t}$ with arclength $S(S_0,t)$, representing the equilibrium configuration of the rod achieved upon removal of the loads. We decompose the total deformation multiplicatively, in terms of the elastic, and growth stretches, respectively $\alpha \eqdef \linepdiff{s}{S}$ and $\gamma \eqdef \linepdiff{S}{S_0}$:
\begin{equation}
    \lambda = \alpha \gamma.\label{eqn:multiplicative_decomp}
\end{equation}
Constitutively, we assume that only $\alpha$ and $\kappa$ generate stresses. Therefore, we posit a quadratic elastic energy density  (energy per unit intermediate length) for a morphoelastic rod with intrinsic curvature $\hat\kappa$, of the form
\begin{align}
    W \pp{\alpha, \kappa} =\frac{1}{2} K \pp{\alpha-1}^2+\frac{1}{2} B(\kappa- \hat\kappa)^2,
\end{align} where $K$ and $B$ are the extensional and bending stiffnesses, respectively. The corresponding constitutive relations are
\begin{subequations}\label{eqn:hooke}
    \begin{align}
 n = \pdiff{W}{\alpha} = K\pp{\alpha - 1},\label{eqn:hooke_resultant}
  \end{align}
  \begin{align}
m = \pdiff{W}{\kappa} = B(\kappa-\hat\kappa),
  \label{eqn:hooke_moment}
 \end{align}
\end{subequations}
  with $ n \eqdef \vec n\cdot\boldsymbol{\tau}$, the longitudinal tension. 
  
There are two remaining important components to model,  the growth of the axonal shaft specified by  $\gamma$, and the remodelling of the curvature $\hat\kappa$. Physically, growth is due to mechanical stresses along the shaft that cause the axon to creep irreversibly above some critical tension \citep{Franze2010,Suter2011,Goriely2015}, while new cytoplasmic material is incorporated simultaneously into the shaft \citep{OToole2011,Oliveri2022b}. During this process, cross-sectional area remains mostly constant, indicating that the cell maintains a homeostatic density \citep{Dennerll1989,Lamoureux1989}; thus we assume that $K$ and $B$ are constant in time and uniform along the rod. To model diffuse growth at each point, we use the exponential growth law  \citep{Oliveri2022a}
\begin{equation}
 \frac{1}{\gamma} \pdiff{\gamma}{t} = \frac{\ramp{n-n_0}}{\eta}, \label{eqn:growth_law}
\end{equation}
 where $\ramp{n-n_0}=n-n_0$ for $n\geq n_0$ ($0$ otherwise); $n_0 \geq 0$ is the \textit{critical tension} below which no growth occurs ; and $\eta$ plays the role of a morphoelastic viscosity (in unit force times time). Henceforth, for simplicity, we neglect the threshold effect and we set $n_0=0$. 

 We also assume a relaxation process for the bending moment due to internal remodelling by using the curvature evolution law
\begin{align}
     \dot {\hat\kappa} = \frac{\kappa  - \hat\kappa}{\tau_\gamma}, 
\end{align}
with a characteristic timescale ${\tau_\gamma}$ of the same order as $\eta/n_0$.

The connection to the mesoscopic model is obtained through the boundary conditions. At the base $S_0=0$, the rod is clamped
\begin{equation}
    \vec r\pp{0} = \vec r_0, \quad \theta \pp{0} = \theta_0,\label{eqn:bcn}
\end{equation}
where $\vec r_0$ and $\theta_0$ are the base position and angle.
At its tip, the rod is  subject to the  traction exerted by the growth cone:
\begin{equation}
m\pp{L_0} = 0, \quad \vec n\pp{L_0} = \vec{F}\pp{\vec r\pp{L_0}, \theta\pp{L_0}},\label{eqn:bcm}
\end{equation}
  where $\vec{F}$ depends on the growth cone position and orientation through the mesoscopic law \eqref{eqn:force-gc-linearised}. 
  
There are three important parameters in the problem, the dimensionless parameter
\begin{align}
    \beta \eqdef \frac{B\zeta}{ K\eta},
\end{align} 
which measures the relative effect of adhesion w.r.t. viscosity, and the two characteristic lengths
\begin{align}
    \Lambda = \sqrt{ \eta / \zeta},\quad \mathcal{L}=\sqrt{ B / F}.
\end{align} 
The \textit{growth length} $\Lambda$ measures the distance over which axonal tension and energy dissipate into the substrate due to friction \citep{OToole2008,Oliveri2021}, and $ \mathcal{L}$ is the \textit{bending length} that measures the typical curvature radius of a beam due to the typical transverse traction $F$ from the growth cone.

At any time, given $\gamma(S_0,t)$ and $\hat\kappa(S_0,t)$, the  boundary-value problem formed by \eqref{eqn:kinematic_relations,eqn:balance,eqn:adhesionforce,eqn:multiplicative_decomp,eqn:growth_law,eqn:hooke} with the boundary conditions \eqref{eqn:bcm,eqn:bcn}, can be solved numerically  to obtain the axon's trajectory. However, the frictional interaction between the axon and its substrate dissipates traction forces over a region close to the growth cone which becomes short,  relatively to the entire axon, as time progresses. Thus, only a short and essentially constant distal section of the axon is dynamic, rapidly causing important computational difficulties due to the Lagrangian specification of our problem. Rather than solving for the system directly, we can take advantage of this effect and focus our attention on  the growth process  localised in a distal region, and freeze the most proximal part of the axon to account for permanent adhesion between the axon and the substrate (and discard that frozen region from the problem). This process is described in the next section.

\subsection{Tip growth approximation\label{tip-growth}}

The existence of a characteristic lengthscale $\Lambda$ can be used to further simplify the problem. For axons, we have $\Lambda\ll\mathcal{L}$ as the force applied by the growth cone only creates small deflections (hence a large radius of curvature $\mathcal L$). This scaling has two important consequences.

First, the effect of the force applied at the tip of the axon by the growth cone on the rest of the axon vanishes quickly away from the tip. It implies that only a short zone at the tip of length $L$ that is of order $\Lambda$ is subject to a force. 
In terms of modelling, we exploit this result computationally by constraining the dynamic region of the axon to a moving active portion of fixed arclength $L$  while segments of the axon found at a greater arclength are fixed and do not need to be updated. In morphoelasticity, this process is referred as \textit{tip growth} \citep{Goriely2017} and is found in  plant shoots or roots, which undergo elongation over an effectively finite apical (or subapical) region  \citep{SILK1979481}.

Second, since the active zone is small with respect to the typical radius of curvature, only small deflections away from the natural state take place in response to the growth cone. The distal end of the beam--the axon's tip--is subject to the locomotory force  $\vec{F}$ produced by the growth cone. The basal end makes the junction between the dynamic region and the frozen region, and is assumed to be clamped from the point of view of the elastic quasi-equilibrium.
Therefore, we can approximate the shape of the active zone by
\begin{equation}\label{eqn:beam-param}
    \vec{r}(\sigma, t) = \vec{\hat{r}}\pp{t}+{\sf x}(\sigma,t)\,\boldsymbol{\hat \tau}(t) + {\sf y}(\sigma, t) \,\boldsymbol{\hat \nu}(t) ,\quad \sigma\in[0,L],
\end{equation}
where $\vec{ \hat r}=\pp{ \hat x ,\hat  y }$ and $\boldsymbol{\hat \tau}(t)$ is the position of the clamp; $
 \boldsymbol{\hat \tau} = \cos \hat \theta \, \vec{e}_x  + \sin \hat \theta \, \vec{e}_y$ and
$
   \boldsymbol{\hat \nu}  = -\sin\hat\theta \, \vec{e}_x \, +  \cos\hat\theta\,\vec{e}_y$  are the tangent and  normal vectors to the curve at the base, with
 $\hat\theta$ the angle of the clamp to the horizontal $\vec e_x$ (henceforth, we denote  quantities evaluated at the base of the beam with a circumflex); $\sf{x}$ and $\sf{y}$, typeset in the {\sf sans-serif} font, denote the coordinates of the rod in the frame of the clamp. For small deflection we have ${\sf x}(\sigma,t) \approx \sigma $, and the rod equations of the previous section simplify to the beam equations \citep[p.~147]{Goriely2017}.

\begin{figure}[ht!]
    \centering
    \includegraphics[width=.8\textwidth]{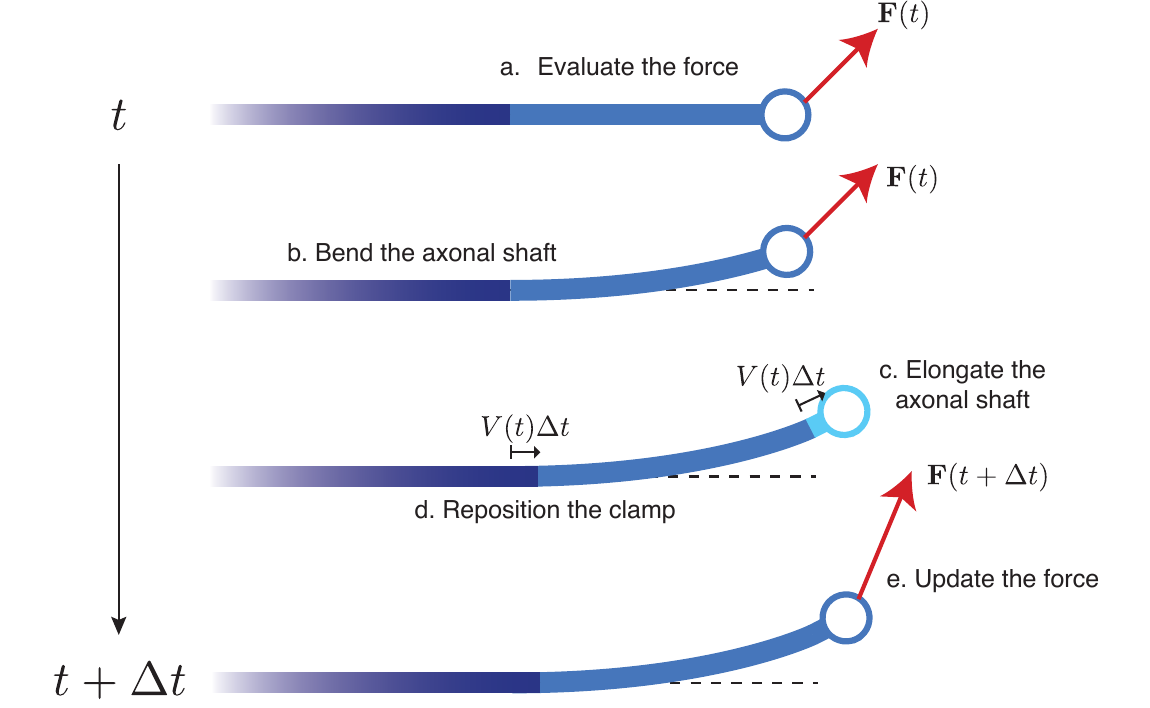}
    \caption{Schematic of the incremental axonal growth process. We assume that only a segment of length $L$ and clamp angle $\hat \theta$ is subject to bending and remodelling. When a force is applied, this clamped beam element bends. Once deformed it grows by an amount $V(t)\Delta t$ and the new clamp is moved and rotated to the new angle $\hat\theta(t)$.}
    \label{fig:remodelling_scheme}
\end{figure}

 There is however an important subtlety appearing in the problem. Since the active zone is moving in time, we have to properly advect the position of the beam element as the axon grows. We keep track of the position and direction of the clamp through the position vector $\vec {\hat r}(t)$ and
model the evolution of $\vec{r}(\sigma, t)$ as the limit in small time step $\Delta t$ of the following process (\cref{fig:remodelling_scheme}): At $t$ the beam is clamped at $\vec{\hat r}(t)$ with direction $\boldsymbol{\hat \tau}(t)$ and is subject to a distal load $\vec{F}(t)$ and viscous friction along its length. First, ${\sf y}(\sigma, t)$ evolves according to the linear beam equations for a time interval of length $\Delta t \ll 1$. Then, a linear segment of length $V(t) \, \Delta t$ is added to the end of the beam in the direction of the tip. We call $V(t)$ the growth rate of the axon. Finally, the length constraint is  enforced by setting the new profile of the beam, $\vec{r}(\sigma, t + \Delta t)$ to be the segment of length $L$ from the tip of the grown, deflected beam profile.

Taking the limit $\Delta t \rightarrow 0$, we obtain the continuous-time evolution equations (see details in \cref{apdx:derivation-pde}):
\begin{subequations}\label{eqn:beam-system}
    \begin{align}&
    \pdiff{{\sf y}}{t} - V(t) \pdiff{{\sf y}}{\sigma} = \frac{1}{\zeta }\pp{\pdiffn{2}{{\sf y}}{\sigma}\vec{F}  \cdot \boldsymbol{\hat \tau}  - B \pdiffn{4}{{\sf y}}{\sigma}} - \sigma \, \widehat{\mathcal{R}}\pp{t},
    \label{eqn:dimensional y equation}\\&
     \pdiff{\hat{\theta}}{t} = \widehat{\mathcal{R}}\pp{t}, \label{eqn:theta hat eqn}\\&
     \pdiff{\vec{\hat r}}{t}  = V(t)\,  \boldsymbol{\hat \tau} ,\label{eqn:dimensional rHat equation}\\&
\widehat{\mathcal{R}}\pp{t} \eqdef V(t) \pdiffn{2}{{\sf y}}{\sigma}(0, t). 
\end{align}
\end{subequations}
The deflection ${\sf y}(\sigma,t)$ satisfies the boundary conditions
\begin{subequations}
\begin{align}
   & {\sf y}(0, t) = 0,\quad 
  \pdiff{{\sf y}}{\sigma}(0, t) = 0, \label{bc: clamp 2}\\&
  \pdiffn{2}{{\sf y}}{\sigma}(L, t) = 0,\quad 
    B \pdiffn{3}{{\sf y}}{\sigma}(L, t) - \vec{F}  \cdot \boldsymbol{\hat \tau} \pdiff{{\sf y}}{\sigma}(L, t) = -\vec{F}  \cdot \boldsymbol{\hat \nu}. \label{bc: force}
\end{align}
\end{subequations}
 Conditions \eqref{bc: clamp 2} define a clamp at $\sigma = 0$, while \eqref{bc: force} enforces the load $\vec F$ applied with no torque on the beam at $\sigma = L$. 

The growth speed $V$ depends on the force exerted by the growth cone on the tip of the axon, and is given, for small deflection and to leading order, by \citep{OToole2008,Oliveri2022a,Oliveri2021}
\begin{equation}
    V  = \frac{\vec{F}  \cdot \vec{t} }{\sqrt{ \eta \zeta}} 
\end{equation}

We have developed a minimal model with two main components. From the microscopic model we have a mapping from substrate rigidity to the traction exerted by the growth cone, and from the mesoscopic model we obtain the shape of the axon as a function of time by locally integrating a clamped beam equation and advecting the clamp accordingly. We are now in a position to use this model to understand the effects of varying substrate rigidity on the motion and shape of axons.

\section{Axonal optics: reflection, refraction and focalisation of axons}

In this section, we explore canonical properties of our theory which capture fundamental emergent properties of the interaction between the axon and a substrate with heterogeneous rigidity. More specifically, we examine two key cases: a straight interface between two regions of uniform stiffness; and a graded field of stiffness defined by a uniform gradient. The question then is to identify the essential laws that govern the motion of an axon within such fields. 

\subsection{Reflection and refraction at a straight stiffness interface}\label{subsection:optics}

We consider a heterogeneous field composed of two regions with different but uniform stiffnesses separated by a straight interface.  In this case, the axon can either grow through the interface (refraction) or bounce against the interface (reflection), in analogy with optic ray theory \citep{Oliveri2021}. 
As the growth cone touches the interface, the force density along the growth cone margin becomes heterogeneous. The total force depends nonlinearly on the position and orientation of the growth cone via \eqref{eqn:gc_resultant_integral}.

\begin{figure}[ht]
\centering
\includegraphics[width=0.9\linewidth]{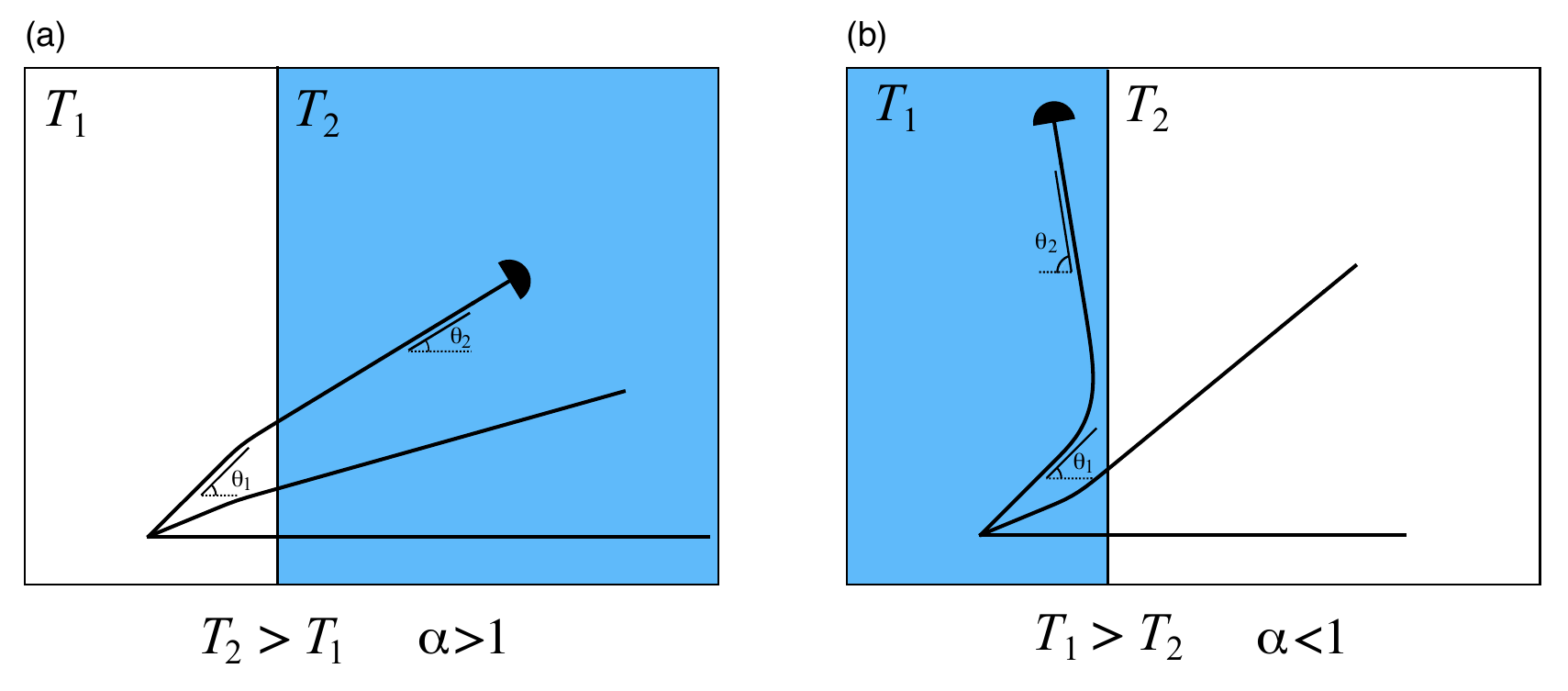}
    \caption{Simulations of axons refracting across and reflecting against an interface with incidence angle $\theta_1$ and reflection or refraction angle $\theta_2$. (a) When the traction exerted by the growth cone is smaller on the right substrate (here $\alpha_T = 5/4$), the axon can only refract.  (b) When the traction exerted by the growth cone is larger on the right substrate (here $\alpha_T = 4/5$), the axon can reflect or refract depending on the incidence angle. Parameters in both scenarios: $\lambda = R/4, \, R = 0.5 L, \, \varphi_0 = \pi/2$, $\beta_a = 0.1$.} 
    \label{fig:reflection refraction section}
\end{figure}

 We assume that the axon grows from the region $x < 0$, as shown in \cref{fig:reflection refraction section}(a), and approaches the boundary with an angle of incidence $\theta_1 \in [0,  {\pi}/{2})$ (the angle between the axon and the normal to the interface). We assume that the interface is sharp but numerically, we approximate this straight interface by 
\begin{equation}
    T\pp{x, y} = \frac{T_1 + T_2}{2} + \frac{T_2 - T_1}{2} \tanh\left( {x}/{\lambda}\right),
\end{equation}
where $T_1 = T\pp{\chi_1}$ and $ T_2 = T\pp{\chi_2}$ are the force densities produced by the growth cone in the left-most and right-most regions, which depend on the stiffness of each region ($\chi_1$, $\chi_2$). The parameter $\lambda$ determines the length scale of transition from $T_1$ to $T_2$ across the interface and is chosen small enough with respect to the growth cone size that the results are not affected by further decreasing it (in practice, we found that $\lambda = R/4$ is suitable).  
Initially, we assume a  clamp position $\vec{r}_0\pp{0}$ far from the boundary in the incidence direction $\theta_1$. As soon as the growth cone leaves the interface region, it follows a straight trajectory and as $t\rightarrow \infty$, we define the \textit{refraction angle} $\theta_2$ with the normal pointing into the right region.  The axon may also reflect against the interface and stay in the left-most region as seen in \cref{fig:reflection refraction section}(b). In this case, the \textit{reflection angle} $\theta_2$ is defined as the angle between the left  normal and the direction of the axon. Let $\theta_{\mathrm{m}} = \lim_{t\rightarrow\infty}\theta_0\pp{t}\mod 2\pi$, then restricting our attention to $\theta_{\mathrm{m}} \in [0, \pi)$ we have
\begin{equation}
    \theta_2 = \begin{cases}
        \theta_{\mathrm{m}}, & \text{if } 0 \leq \theta_{\mathrm{m}} < \pi/2,\\
        \pi - \theta_{\mathrm{m}}, & \text{if } \pi/2 \leq \theta_{\mathrm{m}} < \pi.
    \end{cases}
\end{equation}

By choosing the characteristic force-scale of the system to be $2 T_1\sin {\varphi_0} $, which is equal to the magnitude of the total growth cone force when $\chi = \chi_1$, we obtain the two dimensionless parameters
\begin{equation}
    \alpha_T \eqdef \frac{T_2}{T_1}, \quad \beta_a \eqdef \frac{B\csc\varphi_0}{2 T_1 L^2}.
\end{equation}
 In particular, $\beta_a$ relates to the typical radius of curvature of the axon, that is,  smaller $\beta_a$ generates larger deflections.

We distinguish the two cases $\alpha_T < 1$ and $\alpha_T > 1$. When $\alpha_T > 1$--see \cref{fig:reflection refraction section}(a)--the durotactic force in the right region is greater than in the left region, thus the durotactic force component tends to pull the axon towards the right region. This results in a refractive behaviour with $\theta_{2}$ smaller than $\theta_1$. 

\begin{figure}[ht!]
\centering
\includegraphics[width=0.4\linewidth]{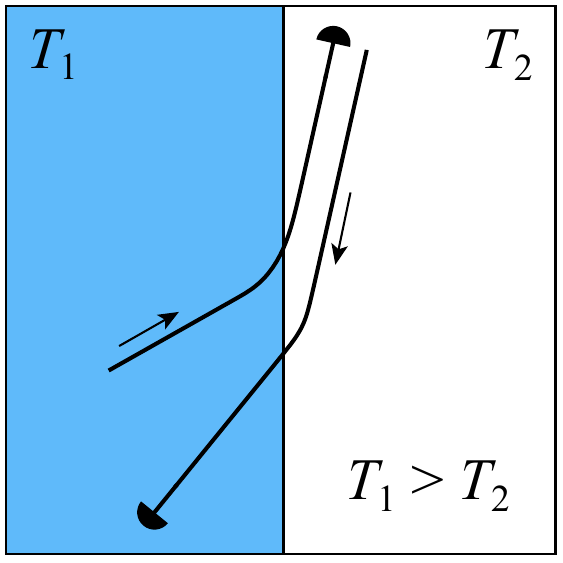}
    \caption{Demonstration of the absence of time-reversibility in the system. Approaching the interface from the left is not the same as approaching from the right. An axon traveling from blue to white with an angle w.r.t the vertical $\theta_1$ ends with an angle $\theta_2$. The same axon traveling from white to blue with an angle $\theta_2$ will end with in a direction given by $\tilde \theta_1\not=\theta_1$. Parameters: $R = 0.5 L, \, \lambda = R/4, \, \varphi_0 = \pi/2$,  $\alpha_T = 4/5,\,\beta_a=0.1$. }
    \label{fig:reflection refraction section2}
\end{figure}

At a superficial level, these observations are reminiscent of the refractive and reflective behaviours of light rays crossing a sharp-interface between two regions of different refractive indices. However, due to the finite size of the growth cone, the interaction with the interface is more complex and there is no simple quantitative analogy with Snell's law. 
Further, a fundamental difference between this model and geometric optics is that we do not have time-reversibility in our system. In  optics, the time-reversal symmetry of Maxwell's equations implies that reflection occurs with $\theta_1 = \theta_2$. As can be seen in \cref{fig:reflection refraction section,fig:reflection refraction section2}, this symmetry does not exist in our model, and in particular, we observe that $\theta_1 \neq \theta_2$ after reflection. In fact, when reflected, axons remain much more parallel to the interface than light rays do.

\subsection{A simple stiffness gradient}
\label{section: graded stiffness field}


Next we examine the case of an axon growing on a substrate with a graded stiffness of the form
\begin{equation}\label{eqn:linear_stiffness_field}
    \chi\pp{x, y} = m_{\chi} y + \chi^*,
\end{equation} 
where the parameter $m_\chi$ is the stiffness gradient.

\begin{figure}[ht!]
    \centering
    \includegraphics[width=.9\textwidth]{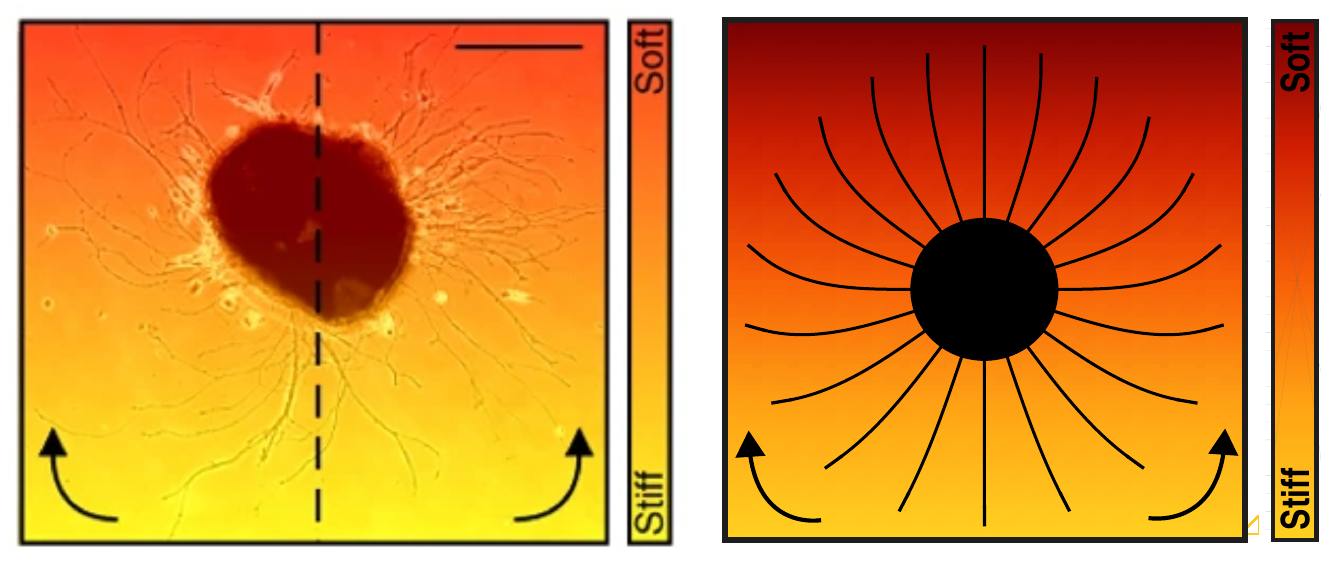}
    \caption{(Left) Axons growing from a Xenopus retina, cultured on a substrate with a linear stiffness gradient. As can be seen, the axons tend to grow more towards the soft side of the substrate \citep[after][]{koser2016mechanosensing}. (Right) Simulation of the rod model, with axons initialised on the boundary of a disk, in the stiffness field \eqref{eqn:linear_stiffness_field}, showing qualitatively similar growth pattern. $m_\chi = -0.1, \,$ $R = 0.5 L, \, \varphi_0 = \pi/2, \, \beta_a=0.1$, $ \beta = f_0 K, \, M = 7.5 f_0 K, \, K = 1/\pp{x^* \E^{x^*}\Gamma\pp{0,x^*}}$.}
    \label{fig:grand-finale}
\end{figure}

Firstly, as a simple illustration of growth on a graded substrate, we initiate multiple axons on the boundary of a disk in a linearly increasing stiffness field, mimicking the experimental setup in \cite{koser2016mechanosensing}. The growth pattern obtained matches qualitatively that seen experimentally, see \cref{fig:grand-finale}, where axons are gradually deflected towards stiffer regions (upper parts of the domain). 

More generally, the local traction exerted by the growth cone exhibits a maximum on the line $\mathcal D$ of optimal stiffness $y = 0$.
We will show that $\mathcal D$ is a locally attracting set for our system in the sense that nearby trajectories are attracted asymptotically to $\mathcal D$. This property is a consequence of the nonmonotonic dependence of the traction force $T$ upon the stiffness $\chi$. 

We consider an axon travelling in a neighbourhood of $\mathcal D$ of typical size $R$. Hence, the growth cone is initially located at the position $\vec r =  r R \vec e_y$, with  $r$ of order one, and travelling nearly parallel to the axis with $\vec t = \vec e_x + \theta \vec e_y + \dots$ with $\theta\ll 1$.  We obtain the  behaviour of this axon by expanding the traction \eqref{eqn:gc_resultant_integral} about $\chi^*$, while treating $R$ and $\theta$ as small parameters:
  \begin{align}
    \rho^{-1} \vec F\pp{\vec r,\theta} =     2R \sin\varphi_0T(\chi^*) \vec t& +   r^2 R^3 \sin\varphi_0T''\pp{\chi^*}m_\chi^2 \vec t + r R^3  \varphi_0 T''(\chi^*)   m_\chi \matr P(\varphi_0)  \nabla \chi \nonumber
      \\&  + \frac{  R^3\varphi_0}{2} T''(\chi^*)\mathbb P\pp{\varphi_0}:\nabla\chi \otimes \nabla\chi + O(R^4),\label{eqn:F-order3}
 \end{align}
 where 
$
     \mathbb P\pp{\varphi_0} \eqdef  \int_{-\varphi_0}^{\varphi_0}     \ephi\otimes \ephi\otimes\ephi \, \pp{{\diff{\varphi}}/{\varphi_0}},
$
 and $\matr P\pp{\varphi_0}$ is defined in \eqref{eqn:definitionP0}.
Equation \eqref{eqn:F-order3}  extends \eqref{eqn:force-gc-linearised} to higher order, which is necessary for our analysis, since the leading-order deflection of $O(R^2)$ vanishes at $\chi=\chi^*$.
Of particular interest is the transverse component $F_n\eqdef \vec F\cdot \vec n$, given by
 \begin{align}
     F_n =
     \frac{1}{6} R^3 \rho T''(\chi^*)m_{\chi }^2 \pp{ 3r  \pp{\varphi_0-\sin \varphi_0 \cos \varphi_0} + 4\theta  \sin ^3\varphi_0
     } +\text{h.o.t.}
 \end{align}
Since $F_n = 0$ when  $\theta=r=0$, there exist trajectories along $\mathcal D$ (however, with a small drift appearing at higher orders and controlled by $T'''(\chi^*)$). 
For $r \gg \theta $, the first term in the parentheses dominates, and, since $T''\pp{\chi^*}<0$, the deflection component has opposite sign of $r$ and grows with $r$ in magnitude. Thus we interpret this term as an elastic-like restoring force which pulls the axon back to the optimal stiffness line $\mathcal D$. Conversely, very close to $\mathcal D$, i.e. for $r \ll  \theta $, the deflection is dominated by the second term and takes now the opposite sign of $\theta$, which, again, will tend to reorient the axon towards the horizontal. We conclude that, close to the line of optimal stiffness, durotaxis has a restoring, stabilising effect which tends to maintain the axon near this line. 

Note, however, that the  longitudinal component $F_t \eqdef \vec F\cdot \vec t \gg F_n$ dominates in the balance of forces, thus there may exist, in principle, conditions for which durotaxis will not suffice to keep the axon in the vicinity of $\mathcal D$. Furthermore, it cannot be directly concluded whether the stable trajectory is attracting (exponential stability), or whether axons will oscillate around it and repeatedly overshoot it.

\begin{figure}[ht!]
    \centering
\includegraphics[width=.7\linewidth]{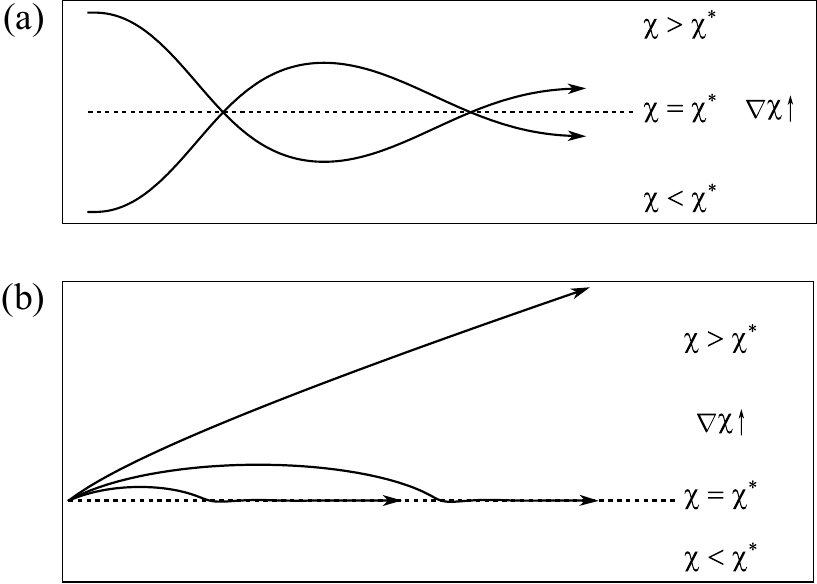}    
    \caption{(a) Axons trajectories converge to the line of optimal stiffness, $\mathcal D$, if sufficiently close to it with sufficiently small inclination. (b) Axons with sufficiently large inclination escape the vicinity of $\mathcal D$.}
    \label{fig:graded stiffness field single axon}
\end{figure}

To further explore the dynamics of the axon near $\mathcal D$, we perform numerical simulations shown in \cref{fig:graded stiffness field single axon}. Firstly, \cref{fig:graded stiffness field single axon}(a) illustrates that, in the vicinity of $\mathcal D$ and for small initial inclination, axons are attracted and converge towards this stable manifold. Second, \cref{fig:graded stiffness field single axon}(b) shows that there may exist  an escape angle above which durotaxis is insufficient to keep the axon closer to $\mathcal D$ despite the negative durotaxis in that region. Clearly, if the axon crosses perpendicularly the line of optimal stiffness, it will not curve back. Whether nearby directions also escape in an infinite domain is neither easy to establish nor directly relevant to the problem. Indeed, in a much larger domain, it is possible that such trajectories curve back to the line of optimal stiffness. However, in practice such axons have migrated away from the region of optimal stiffness and are subject to other stimuli.

In conclusion these examples illustrate that, even for a simple uniform gradient, the nonmonotonic, nonlinear dependence of the locomotory force upon stiffness implies the existence of preferential regions where axons are expected to converge. Such a mechanism may be key to understanding neuronal focalisation towards regions of specific stiffness.

\section{Biological applications}


Taken together, these findings highlight two primary modes of guidance: deflective guidance (via reflection and refraction) and attractive guidance (via emergent attraction to regions of optimal stiffness). Lastly, we illustrate the potential role of these guidance modes in a biological scenario.

During early neurodevelopment, Xenopus retinal ganglion cells (RGCs)  establish connections with the brain as their axons exit the retina through the optic nerve and migrate towards the brain \citep{koser2016mechanosensing,Thompson2019}. These axons grow collectively, cross the midline at the optic chiasm, and then travel along the contralateral brain surface towards the optic tectum, where they terminate. Towards the end of this journey, a postero-anterior gradient in brain tissue rigidity emerges, which strongly correlates with the stereotypical caudal turn observed in RGC axons as they traverse the mid-diencephalon en route to the tectum \citep{koser2016mechanosensing}.

\cite{Oliveri2021} investigated the influence of this stiffness gradient on the deflection of an entire bundle of axons, under the assumption that the axons are mechanically coupled. In their model, differential stiffness sensing generates a velocity differential across the bundle's leading front, which explains its coordinated deflection. However, in biological systems, the mechanical coupling between axons is mediated by cell-cell adhesion, which is highly regulated and may be more or less loose  \citep{Smit2017}.

In this study, we address the problem from the opposite perspective by assuming the absence of mechanical coupling between axons in the bundle, which is compatible with a scenario where axons follow a leader cell \citep{Hentschel1999}. Using our model, we focus on the trajectories of individual axons, offering a complementary understanding of the role of stiffness gradients in axonal guidance.


\begin{figure}[ht!]
    \centering
\includegraphics[width=.99\linewidth]{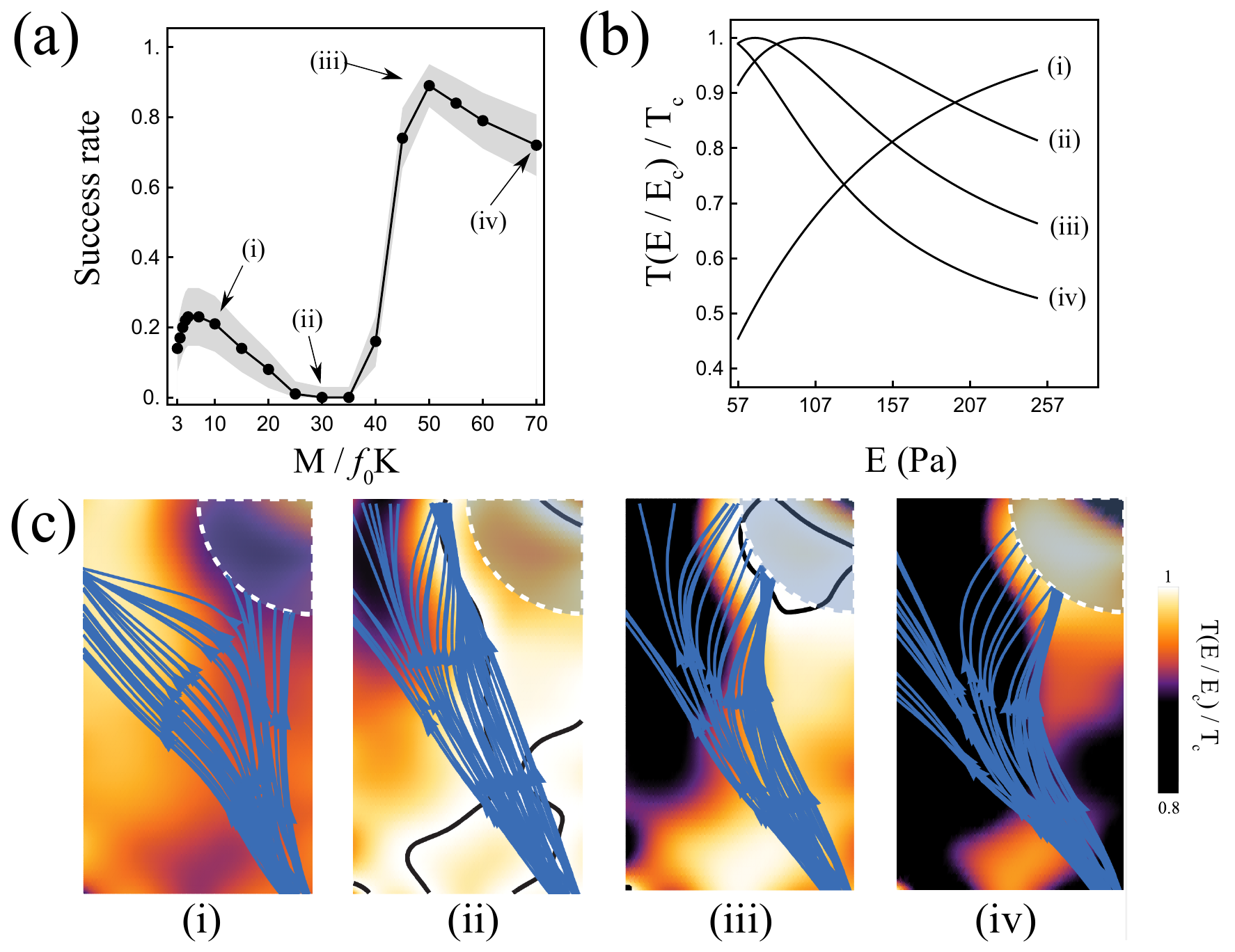}    
    \caption{(a) Success rate as a function of $M$. (b) Traction as a function of $E$ for the values of $M$ indicated in (a). (c) Density map of traction for the values of $M$ indicated in (a) overlaid with representative trajectories of axons for the given parameters. $K = 0.1, \,\beta = f_0 K , \, T_c = 2 \sin\pp{\varphi_0} f_0 K \mathcal{H}\pp{x^*}, \,E_c = 6000$ \pascal, $R = 0.5 L, \, \varphi_0 = \pi/2$,  $\beta_a=0.1$.}
    \label{fig:optic tract}
\end{figure}

To study the trajectory of single axons in the optic tract we use a brain rigidity field, $E\pp{x,y}$ with units of pressure, obtained by atomic force microscopy \citep{Thompson2019}. We investigate the trajectories of axons growing on a substrate with stiffness field $\chi\pp{x, y} = E\pp{x,y} / E_c$, where $E_c$ is a characteristic pressure. We simulated $N = 100$ axons, varying their initial position and direction around the entry point of the domain. The target zone (optic tectum) was defined as the quadrant seen in \cref{fig:optic tract}(c) and for each set of parameters we record the success rate $N_s / N$, characterising the ratio of axons successfully reaching the target (with number $N_s$) to the total number $N$.

In previous sections, we  showed that the key features of the traction-stiffness curve $T\pp{\chi}$ relating to axon guidance are $\chi^*$, $T''\pp{\chi^*}$, and $T_\infty$. These features can be directly controlled through myosin contractility $M$ (see \cref{subsection:contractility}) which can be targetted through pharmacological treatments in experiments such as application of the myosin inhibitor blebbistatin \citep{kovacs2004mechanism}, or by modulating the number of myosin molecules \citep{bangasser2017shifting,bangasser2013determinants}; or via the clutch properties \citep{bangasser2017shifting,bangasser2013determinants}.

Here we investigated the effect of myosin contractility  on the success rate and found that  success rate is a nonmonotonic function of $M$. The analysis of the success rate as a function of contractility  allows us to identify  four regimes depending on  $M$, where the success rate is in turn decreasing, constant, increasing and decreasing again; see  \cref{fig:optic tract}(a).

We can understand this nonmonotonic behaviour by looking at the resulting traction field $T\pp{E\pp{x,y}/E_c}$ in the four different regimes. We investigate a representative case from each regime; (i)--(iv) in \cref{fig:optic tract}(a). In the domain, the rigidity ranges from $57$ \pascal{} to $250$ \pascal. For $M / f_0 K = 10$, the optimal stiffness is greater than $250$ \pascal. Therefore, traction is an increasing function of rigidity between $57$ \pascal{} and $250$ \pascal; see (i) in \cref{fig:optic tract}(b). Since the left side of the domain is stiffer than the right side, the traction $T\pp{E\pp{x,y}/E_c}$ smoothly increases from right to left. As a result, most axons are attracted to the left side of the domain leading to a poor success rate of $21\%$. 

As $M / f_0 K$ increases further, the optimal stiffness decreases and enters the range of stiffness values present in the domain; see (ii) and (iii) in \cref{fig:optic tract}(b). As a result, contours of optimal stiffness form within the domain and axons tend to follow them. In the case of $M / f_0 K = 35$, one of these contours leads most of the axons away from the target leading to a catastrophic success rate of $0\%$; see \cref{fig:optic tract}(c), (ii). 

For greater values of $M / f_0 K$, with the optimal stiffness decreasing further and the decay of $T\pp{E/E_c}$ to $T_\infty$ becoming faster, the domain becomes roughly bisected by an interface with a region of high traction on the right side of the domain and a region of low traction on the left side of the domain. This situation is similar to the example studied in \cref{subsection:optics} where axons can reflect against a sharp interface between regions of different rigidities. Indeed, we see that for $M / f_0 K = 50$, most axons reflect against a well-formed interface between the two regions of low and high traction, and make it to the target; see \cref{fig:optic tract}(c), (iii). Because the interface is not perfectly straight, many axons that are not initially reflected by their interaction with the interface, re-approach the interface later on. In this case, the axons are refracted into the region of high traction and subsequently reach the target. This combination of reflection and refraction results in the high success rate of $89\%$. For $M/f_0 K > 50$, a combination of reflection and refraction again results in high success rate. However, the placement of the interface is less favourable and fewer axons reach the target; see \cref{fig:optic tract} (c), (iv).

These simulations suggest that durotaxis may play a role in guiding the optic tract to the tectum. However, obtaining accurate parameter estimates remains a significant challenge, making it difficult to draw definitive conclusions about the precise role of durotaxis in this process. Additionally, collective effects and chemotaxis are well-established as dominant mechanisms in optic tract guidance.

Nonetheless, the results illustrate how paradigms such as reflection, refraction, and the tracking of lines of optimal stiffness can, in principle, guide axons in biologically realistic scenarios. Furthermore, the findings underscore the critical role  of the traction stiffness curve $T\pp{\chi}$ in controlling the dynamics. As emphasised in this paper, axonal trajectories in durotaxis are dictated by the stiffness field and the three key properties of the traction stiffness curve: $\chi^*$, $T''\pp{\chi^*}$, and $T_\infty$. The choice of these properties (e.g. via the parameter $M$) control the outcomes. Our results suggest that, under the proposed model of durotaxis, the most plausible mechanism for guiding axons to the optic tract involves a combination of reflection and refraction, rather than following lines of optimal stiffness.


\section{Discussion}

Axons are examples of active filaments that process multiple and complex stimuli and dynamically adapt their growth to these stimuli, a process similar to tropic phenomena in plants \citep{moulton2020multiscale} or the response of soft liquid crystal elastomer rods \citep{goriely2023rod}. 
Therefore, mechanically, axons are slender filaments influenced by their environment, which affects their growth, deformation, and motion. The problem is then to model the evolution of such a structure. 

Here we adopted a multiscale approach to model the migration of an axon, combining (i) the molecular interaction between an axon and its substrate; (ii) the integration of such interactions over the growth cone to obtain the forces acting at the tip of the axon;  (iii) modelling the growing axon as a morphoelastic rod in response to both stimuli and interaction with its environment; and (iv) simplifying this model for computational purposes based on the fact that growth takes place close the growth cone.


In contrast with many models for axon guidance based on Brownian-type dynamics \citep[see][and references therein]{maskery2005deterministic,Oliveri2022a}, growth is modelled  explicitly, as well as the mechanics of axonal shaft bending and elongation, and the mechanics of cell adhesion underlying locomotion.  
The primary challenge in constructing a mechanistic model of axonal guidance lies in simultaneously accounting for axonal mechanics, growth cone mechanics, and actomyosin dynamics. To address this problem, we adopt a multiscale, mean-field framework. Following \cite{Chan2008} and \cite{sens2013rigidity}, we begin by modelling the fundamental interaction between a single actin filament and the substrate. This local interaction law is then employed in a second step to derive the total force exerted across the growth cone.

Crucially, we do not model explicitly the entire actin-myosin lamellipodium by taking into account each individual actin filament, but, instead, we derive a minimal expression for the growth cone traction in a shallow stiffness gradient. This approach contrasts with models such as \cite{Aeschlimann2000}, where individual filaments are represented explicitly in a computational framework. By integrating this growth cone force with a reduced beam model, which accounts for the gradual fixation of the axon to its substrate and incorporates a finite distal growth zone, we obtain a simple tractable model that can be easily implemented.

This approach reveals key guiding principles of durotaxis. First, the mechanical interaction between the axon and the substrate, governed by Bell-type adhesion kinetics, produces a highly nonlinear and potentially nonmonotonic dependence of the force $T$ generated by an actin filament on substrate stiffness. Such previously-identified nonmonotonicity \citep{Chan2008,sens2013rigidity,bangasser2013master} arises from a phenomenon known as the \textit{stick-slip instability}, where excessive stiffness leads to rapid breakage of cross-linkers, reducing the generated force compared to more moderate stiffness levels. Critically, this nonmonotonic behavior of $T$ is essential for explaining the coexistence of positive and negative durotactic regimes, which can arise without the need for separate cellular mechanisms.

At the cellular scale, this behaviour implies the existence of stable, attractive regions with optimal stiffness where axons are likely to converge due to enhanced adhesion. Notably, the idea that a simple stiffness gradient can create distinct regions via nonlinear cellular sensing strongly parallels patterning mechanisms such as Wolpert’s \textit{French flag} model \citep{wolpert2017french}. This analogy may hold particular significance for understanding mechanically induced neuronal segmentation, as observed by \cite{Schaeffer2022}.

Second, inspired by the analogy proposed by \cite{Oliveri2021}, we demonstrate that axons encountering a straight stiffness interface may exhibit behaviours analogous to optical refraction or reflection, depending on the incidence angle and mechanical conditions. This observation raises the intriguing possibility of interpreting durotactic guidance through mechanisms analogous to optical lensing or channelling, albeit governed by laws distinct from Snell's law. This opens an exciting avenue for future research to uncover the broader principles of ``axonal optics'' that emerge from this model.

Axonal growth is a highly complex cellular process, influenced by numerous interdependent signals and growth conditions, and is inherently noisy. In such a context, the prospect of establishing simple, universal laws that capture the detailed intricacies of axonal guidance may appear unattainable. Nevertheless, by employing a minimal set of assumptions grounded in first principles and biological observations, we uncovered fundamental qualitative principles that govern durotaxis.

Our multiscale framework is a general theory that integrates microscale interactions between axons and substrate with macroscale motion. It is presented here in the simplest setting and, as such, is not designed to replicate particular experiments. Yet, achieving more quantitative applications is possible but will require extending the model to account for additional factors, such as chemotactic influences—particularly the poorly understood interplay between chemotactic sensitivity and mechanics \citep{Franze2020}; steric interactions with other cells; the mechanical characteristics and particularities of different substrates in both two and three dimensions \citep{SANTOS2020107907}; more intricate axonal behaviours, such as contraction and collapse \citep{Recho2016}; stochastic effects; domain curvature and its influence on cell growth \citep{smeal2005substrate}; and collective chemical and mechanical phenomena \citep{chaudhuri2011model,de2007collective,Oliveri2021,Hentschel1999}.

Despite these phenomenological limitations, the proposed model provides a foundational framework and a key step towards a comprehensive, multiscale and multiphysics field theory of axonal development that can be  scaled up to the entire brain.

\section*{Acknowledgements}

C.K. acknowledges funding through a studentship from the Engineering and Physical Sciences Research Council (EPSRC) under project reference 2580825.
The authors thank Kristian Franze for insightful discussions.

\appendix

\section{Derivation of the beam model\label{apdx:derivation-pde}}

Here we derive the evolution laws for the dynamic section modelled by an inextensible growing beam (\cref{tip-growth}). The dynamic distal section of the axon is represented by a spatial curve $\vec r(\sigma, t)$ given as
\begin{equation}\label{eqn:defr}
    \vec{r}(\sigma, t) = \vec{\hat{r}}\pp{t}+{\sf x}(\sigma,t)\,\boldsymbol{\hat \tau}(t) + {\sf y}(\sigma, t) \,\boldsymbol{\hat \nu}(t)
\end{equation}
Here, $\sigma \in \bb{0,L}$ measures the arclength from the base of the dynamic section; and where $\vec{\hat{r}}$,  $\boldsymbol{\hat \tau}$ and $\boldsymbol{\hat \nu}$ indicate respectively the position, and the tangent and normal vectors to the curve at $\sigma=0$. The coordinates $\sf x$ and $\sf y$ indicate the position of the curve in the local frame $\left\{\boldsymbol{\hat \tau},\boldsymbol{\hat \nu} \right\}$. 
These definitions imply the boundary conditions
\begin{equation}
    {\sf x}\pp{0, t} = {\sf y}\pp{0, t} = 0,\quad  \pdiff{{\sf x}}{\sigma}\pp{0, t} = 1, \quad \pdiff{{\sf y}}{\sigma}\pp{0, t} = 0.
    \label{eqn:bc1}
\end{equation}
Further, ${\sf x}, {\sf y}$ can be obtained by integrating the system
\begin{equation}
    \pdiff{\sf x}{\sigma} = \cos\theta, \quad\pdiff{\sf y}{\sigma} = \sin\theta,
\end{equation}
where $\theta\pp{\sigma,t}$ indicates the polar angle of the tangent $\partial{\vec r}/\partial{\sigma}$ at arclength $\sigma$ and time $t$, in the local frame $\left\{\boldsymbol{\hat \tau},\boldsymbol{\hat \nu} \right\}$. That is,
\begin{equation}
    \pdiff{\vec r}{\sigma} = \cos\theta \,\boldsymbol{\hat \tau} + \sin\theta \,\boldsymbol{\hat \nu}.
\end{equation}

During a small time increment $\Delta T $ we update the shape of the axon by first evolving the curve according to the overdamped beam equations \eqref{eqn: overdamped}, assuming that the axon is clamped at $\hat{\vec{r}}\pp{t}$ with angle $\hat{\theta}\pp{t}$, and subject to the force $\vec{F}\pp{t} $ applied at $\sigma = L$ with zero torque. We introduce the auxiliary function
\begin{equation}
    \tilde{\vec{r}}\pp{\sigma, T, t} = \vec{\hat r}\pp{t}+ \tilde{\sf x}\pp{\sigma, T, t} \, \boldsymbol{\hat \tau} + \tilde{\sf y}\pp{\sigma, T, t} \boldsymbol{\hat \nu}.
\end{equation}
Here, $\tilde{\vec{r}}\pp{\sigma, T, t}$ describes the evolution of the axon profile, in time $T$, with initial condition $\vec{r}\pp{\sigma, t}$ and clamp angle $\hat{\theta}\pp{t}$, subjected to constant force $\vec{F}\pp{t}$. The functions $\tilde{\sf x}\pp{\sigma, T, t}$ and $\tilde{\sf y}\pp{\sigma, T, t}$ satisfy the overdamped beam equations in the frame $\{\vec{\boldsymbol{\hat \tau}}, \vec{\boldsymbol{\hat \nu}} \}$, with spatial variable $\sigma$ and time variable $T$.
\begin{subequations}\label{eqn: overdamped}
\begin{equation}\label{eqn:balance-forces-tilde}
    B \pdiffn{2}{\tilde{\theta}}{\sigma} + \tilde{\sf n}_y \cos\tilde{\theta} - \tilde{\sf n}_x \sin\tilde{\theta} = 0,\end{equation}
    \begin{equation}
    \pdiff{\tilde{\sf n}_x}{\sigma} = \zeta \pdiff{\tilde{\sf x}}{T},  \quad  \pdiff{\tilde{\sf n}_y}{\sigma} = \zeta \pdiff{\tilde{\sf y}}{T},\end{equation}\begin{equation}\label{eqn:kinematics-tilde-xy}
    \pdiff{\tilde{\sf x}}{\sigma} = \cos  {\tilde{\theta}}, \quad \pdiff{\tilde{\sf y}}{\sigma} = \sin {\tilde{\theta}},
\end{equation}
\end{subequations}
where $\tilde{\sf n}_x$ and $\tilde{\sf n}_y$ are the components of the internal resultant force $\boldsymbol{\tilde{\sf n}} = \tilde{\sf n}_x \boldsymbol{\hat \tau} + \tilde{\sf n}_y \boldsymbol{\hat \nu}$ at time $T$. 
We enforce the boundary and initial conditions 
\begin{subequations} \label{eqn:overdamped bcs}
    \begin{equation} \label{eqn:nxny}
    \tilde{\sf n}_x\pp{L, T, t} = \vec{F}\pp{t} \cdot \boldsymbol{\hat \tau}(t),  \quad \tilde{\sf n}_y\pp{L, T, t} = \vec{F}\pp{t} \cdot \boldsymbol{\hat \nu}(t),   \end{equation}
    \begin{equation}
     \tilde{\theta}\pp{0, T, t} = 0,  \quad  \pdiff{\tilde{\theta}}{\sigma}\pp{L, T, t} = 0,\label{eqn: bc theta aux}    \end{equation}
     \begin{equation}
     \tilde{\sf x}\pp{0, T, t} = 0,  \quad \tilde{\sf y}\pp{0, T, t} = 0, \label{eqn: bc y aux} 
     \end{equation}
     \begin{equation}
    \tilde{\sf x}\pp{\sigma,0, t} = {\sf x}\pp{\sigma, t}, \quad \tilde{\sf y}\pp{\sigma,0, t} = {\sf y}\pp{\sigma, t}.
\end{equation}
\end{subequations}

From here, we can update the axon's shape according to \eqref{eqn: overdamped} integrated over a small time increment $\Delta T$. That is, we define an intermediate curve
\begin{equation}
\vec{\tilde{r}}_\mathrm{inter}\pp{\sigma} = \vec{\tilde{r}}\pp{\sigma, \Delta T, t}.
\end{equation}
The updated axon after growth is then obtained by advancing the clamp position along this intermediate profile by a small length $\Delta \sigma = V\pp{t} \Delta T$,
\begin{equation}
    \vec{r}\pp{\sigma, t + \Delta T} = \vec{\tilde{r}}_\mathrm{inter}\pp{\sigma + \Delta \sigma}.
\end{equation}
Expanding the right-hand side to second order in $\Delta T$, rearranging the terms and taking the limit $\Delta T \rightarrow 0$, we obtain
\begin{align}
    \pdiff{\vec{r}}{t}\pp{\sigma, t} =& \pp{V\pp{t} \pdiff{\tilde{\sf x}}{\sigma}\pp{\sigma, 0, t} + \pdiff{\tilde{\sf x}}{T}\pp{\sigma, 0, t}} \boldsymbol{\hat \tau} \nonumber\\&
    + \pp{V\pp{t} \pdiff{\tilde{\sf y}}{\sigma}\pp{\sigma, 0, t} + \pdiff{\tilde{\sf y}}{T}\pp{\sigma, 0, t}} \boldsymbol{\hat \nu}.
    \label{eqn:evolution 1}
\end{align}
Differentiating \eqref{eqn:defr} we derive
\begin{equation}
    \pdiff{\vec{r}}{t}\pp{\sigma, t} = \fdiff{\hat{\vec{r}}}{t}+ \pp{\pdiff{{\sf x}}{t} - {\sf y} \fdiff{\hat \theta}{t}} \vec{\boldsymbol{\hat{\tau}}} + \pp{\pdiff{{\sf y}}{t} + {\sf x} \pdiff{{\hat \theta}}{t}} \vec{\boldsymbol{\hat{\nu}}} ,
    \label{eqn:evolution 2}
\end{equation}
where ${\hat \theta}$ is the angle between $\boldsymbol{\hat \tau}$ and the horizontal direction.
Equating the right-hand sides of \eqref{eqn:evolution 1,eqn:evolution 2} we obtain
\begin{subequations}
    \begin{equation}
    \pdiff{{\sf x}}{t} -{\sf y} \fdiff{\hat{\theta}}{t} + \pdiff{\hat{\vec{r}}}{t} \cdot \boldsymbol{\hat \tau} = V\pp{t} \pdiff{\tilde{\sf x}}{\sigma}\pp{\sigma, 0, t} + \pdiff{\tilde{\sf x}}{T}\pp{\sigma, 0, t},
\end{equation}
\begin{equation}
    \pdiff{{\sf y}}{t} + {\sf x} \fdiff{\hat{\theta}}{t} + \pdiff{\hat{\vec{r}}}{t} \cdot \boldsymbol{\hat \nu} = V\pp{t} \pdiff{\tilde{\sf y}}{\sigma}\pp{\sigma, 0, t} + \pdiff{\tilde{\sf y}}{T}\pp{\sigma, 0, t}.
\end{equation}
\end{subequations}
To obtain the evolution equation for $\hat{\vec{r}}$ we set $\sigma = 0$ in the above equations. We note that due to \eqref{eqn:bc1} we have $\linepdiff{{\sf x}}{t}\pp{0, t} = \linepdiff{{\sf y}}{t}\pp{0, t} = 0 $ and so
\begin{equation}
    \pdiff{\hat{\vec{r}}}{t} = V\pp{t} \boldsymbol{\hat \tau}
    \label{eqn: r hat}
\end{equation}
We derive the evolution equation for $\hat{\theta}$ by imposing (to first order in $\Delta T$) that
\begin{equation}
    \pdiff{\vec r}{\sigma}\pp{0, t + \Delta T} = \pdiff{\vec r}{\sigma}\pp{V(t) \Delta T, t}.
\end{equation}
Expanding the left-hand side in $t$ and the right-hand side in $\sigma$ to second order in $\Delta T$, rearranging and taking the limit $\Delta T \rightarrow 0$ gives
\begin{equation}
    \fdiff{}{ t}\pp{\pdiff{\vec r}{\sigma}\pp{0, t}} = V\pp{t} \pdiffn{2}{\vec r}{\sigma}\pp{0, t},
\end{equation}
which simplifies to
\begin{align}
    & \pdiffn{2}{{\sf x}}{\sigma}\pp{0, t} = 0, \\
    &\fdiff{\hat{\theta}}{t} = V\pp{t} \pdiffn{2}{{\sf y}}{\sigma}\pp{0, t}. \label{eqn: theta hat} 
\end{align}
Therefore the governing equations become
\begin{subequations}\label{eqn:full}
    \begin{align}
    \pdiff{{\sf x}}{t} - {\sf y} V\pp{t} \pdiffn{2}{{\sf y}}{\sigma}\pp{0, t} +  V\pp{t} &= V\pp{t} \pdiff{\tilde{\sf x}}{\sigma}\pp{\sigma, 0, t} + \pdiff{\tilde{\sf x}}{T}\pp{\sigma, 0, t}, \label{eqn:full 1}\\
    \pdiff{{\sf y}}{t} + {\sf x} V\pp{t} \pdiffn{2}{{\sf y}}{\sigma}\pp{0, t} &= V\pp{t} \pdiff{\tilde{\sf y}}{\sigma}\pp{\sigma, 0, t} + \pdiff{\tilde{\sf y}}{T}\pp{\sigma, 0, t}. \label{eqn:full 2}
\end{align}
\end{subequations}
Next, we assume that the normal component of the force $\vec F$ with respect to the clamp axis is small. That is,
\begin{equation}
    \vec F \cdot \boldsymbol{\hat \nu} = \epsilon N_y,
\end{equation}
for $\epsilon \ll 1$ and where $
    \vec F \cdot \boldsymbol{\hat \tau} $ and $  N_y $ are $ \mathcal{O}(1)$.
Consequently, the axon is subject to small deflections from the axis spanned by $\boldsymbol{\hat{\tau}}$. i.e.
\begin{equation}\label{eqn:xytheta approx}
    {\sf x} = \sigma + O\pp{\epsilon^2},\quad {\sf y} = \epsilon Y\pp{\sigma,t} + O\pp{\epsilon^2},\quad \theta = \epsilon \Theta\pp{\sigma,t} + O\pp{\epsilon^2},
\end{equation}
for $\epsilon \ll 1$.
Plugging \eqref{eqn:xytheta approx} into \eqref{eqn:full} we obtain
\begin{subequations}
\begin{align}
   \pdiff{\tilde{\sf x}}{T}\pp{\sigma, 0, t} &= O\pp{\epsilon^2} , \\
  \pdiff{\tilde{\sf y}}{T}\pp{\sigma, 0, t}-  \epsilon \pp{\dot{Y} + \sigma V\pp{t} \pdiffn{2}{Y}{\sigma}\pp{0, t} -V\pp{t} \pdiff{Y}{\sigma}}  &= O\pp{\epsilon^2}.
\end{align}
\end{subequations}
For the next step, it is important to note that since we have $\partial{{\sf y}}/\partial{\sigma} = \sin  \theta$ and $\tilde{y}\pp{\sigma,0,t} = y\pp{\sigma, t},$ it follows that $\partial{\tilde{{\sf y}}}/\partial{\sigma}\pp{\sigma,0,t}=\partial{{\sf y}}/\partial{\sigma}$ and therefore
\begin{equation}
    \tilde{\theta}\pp{\sigma,0,t} = \theta\pp{\sigma,t}.
\end{equation}
We can now obtain an expression for $\linepdiff{\tilde{\sf y}}{T}\pp{\sigma, 0, t}$ to first order in $\epsilon$. Positing
\begin{equation}
    \tilde{\theta}\pp{\sigma,0, t} = \theta\pp{\sigma,t}=\epsilon \Theta\pp{\sigma, t} + O\pp{\epsilon^2}.
\end{equation}
with $\Theta$ a function of order unit,
using \eqref{eqn:kinematics-tilde-xy,eqn:balance-forces-tilde}, and differentiating with respect to $\sigma$, we obtain 
we obtain
\begin{equation}
    \epsilon B \, \pdiffn{3}{\Theta}{\sigma} + \pdiff{\tilde{\sf n}_y}{\sigma} - \tilde{\sf n}_x \pdiffn{2}{\tilde{\sf y}}{\sigma}\pp{\sigma,0;t} =O\pp{\epsilon^2} ,
\end{equation}
which is equivalent to the beam equation
\begin{equation}
    \epsilon B \, \pdiffn{4}{Y}{\sigma} + \zeta \pdiff{\tilde{\sf y}}{T} - \epsilon \tilde{\sf n}_x \pdiffn{2}{Y}{\sigma} =O\pp{\epsilon^2} .
\end{equation}
Thus,
\begin{equation}
    \pdiff{\tilde{\sf y}}{T}\pp{\sigma,0;t} = \frac{\epsilon}{\zeta} \pp{\tilde{\sf n}_x \pdiffn{2}{Y}{\sigma} - B \pdiffn{4}{Y}{\sigma}} + O(\epsilon^2).
\end{equation}
To leading order we thus have
\begin{equation}
    \pdiff{Y}{t} - V\pp{t} \pdiff{Y}{\sigma} = \frac{1}{\zeta}\pp{\vec{F}\pp{t} \cdot \boldsymbol{\hat \tau} \pdiffn{2}{Y}{\sigma} - B \pdiffn{4}{Y}{\sigma}} - \sigma V\pp{t} \pdiffn{2}{Y}{\sigma}\pp{0, t}.
    \label{eqn: linear Y}
\end{equation}
We obtain the first three boundary conditions for $Y$ by evaluating \eqref{eqn: bc theta aux,eqn: bc y aux} at $T = 0$, to first order in $\epsilon$, giving
\begin{equation}
    Y\pp{0, t} = 0, \quad \pdiff{Y}{\sigma}\pp{0, t} = 0, \quad \pdiffn{2}{Y}{\sigma}\pp{L, t} = 0.
    \label{eqn: Y bc 1}
\end{equation}
The fourth boundary condition for $Y$ is obtained similarly, by evaluating \eqref{eqn:balance-forces-tilde} at $\sigma = L$ and $T = 0$, to $O(\epsilon)$,
\begin{equation}
     B \pdiffn{3}{Y}{\sigma}(L, t) - \vec{F} \cdot \boldsymbol{\hat \tau} \pdiff{Y}{\sigma}(L, t) = -N_y.
     \label{eqn: Y bc 2}
\end{equation}
Equation \eqref{eqn: linear Y} along with boundary conditions \eqref{eqn: Y bc 1,eqn: Y bc 2} constitute the linearised system for $Y$ under small deflections. Multiplying each of these equations by $\epsilon$, we obtain the linearised equation for $\sf y$, to first order in $\epsilon$. The full system is then
\begin{align}
    &\pdiff{\hat{\vec{r}}}{t} = V\pp{t} \boldsymbol{\hat \tau},\\
    &\pdiff{\hat{\theta}}{t} = V\pp{t} \pdiffn{2}{{\sf y}}{\sigma}\pp{0, t},\\
    &\pdiff{\sf y}{t} - V\pp{t} \pdiff{\sf y}{\sigma} = \frac{1}{\zeta}\pp{\vec{F}\pp{t} \cdot \boldsymbol{\hat \tau} \pdiffn{2}{\sf y}{\sigma} - B \pdiffn{4}{\sf y}{\sigma}} - \sigma V\pp{t} \pdiffn{2}{\sf y}{\sigma}\pp{0, t},\\
    &\sf y\pp{0, t} = 0, \quad \pdiff{\sf y}{\sigma}\pp{0, t} = 0, \quad \pdiffn{2}{\sf y}{\sigma}\pp{L, t} = 0,\\
    &B \pdiffn{3}{\sf y}{\sigma}(L, t) - \vec{F} \cdot \boldsymbol{\hat \tau} \pdiff{\sf y}{\sigma}(L, t) = -\vec{F} \cdot \boldsymbol{\hat \nu}.
\end{align}
These equations make up the system \eqref{eqn:beam-system} used in the main text. 

\section{Numerics and implementation details\label{apdx:numerics}}
To solve the system for $(\boldsymbol{\hat{r}}, \hat \theta, {\sf y})$ we spatially discretise \eqref{eqn:dimensional y equation} using a spectral method. This approach  approximates the derivatives of a given function by the derivatives of a Chebyshev polynomial interpolation of that function on a discrete grid \citep{trefethen2000spectral,Press2007}.  
To that end, we first transform the spatial domain to the interval $[-1,1]$. We start by rewriting \eqref{eqn:dimensional y equation} in terms of the dimensionless variable
\begin{equation}\label{eqn:change-variable-sigma-X}
    X =   {2 \sigma}/{L} - 1 \in \bb{-1,1}.
\end{equation}
This allows us to discretise the spatial domain according to the Chebyshev grid \citep{trefethen2000spectral}
\begin{equation}
    X_j = \cos\pp{j \pi/N}, \quad j = 0, \dots, N.
\end{equation}
Thus, we can convert the PDE \eqref{eqn:dimensional y equation} into a system of ordinary differential equations. First, we define
\begin{equation}
    {\sf y}_j\pp{t} = {\sf y}\pp{X_j, t},
\end{equation}
and the vector
\begin{equation}
    \boldsymbol{{\sf y}} = \pp{{\sf y}_0, \dots, {\sf y}_N}.
\end{equation}
The spatial derivatives of ${\sf y}(X, t)$ evaluated on the Chebyshev grid may be approximated by multiplying $ \boldsymbol{{\sf y}}$ with the Chebyshev differentiation matrix $D_N$ \citep{trefethen2000spectral}. e.g.
\begin{equation}
    \pdiffn{k}{{\sf y}}{X}\pp{X_j, t} \approx \left[D_N^k \boldsymbol{\sf y}\right]_j.
    \label{eqn: derivative approximation}
\end{equation}
By replacing instances of the spatial derivatives of ${\sf y}(X, t)$ with the above approximation in the rescaled equation, we obtain an ordinary differential equation for $\boldsymbol{\sf y}\pp{t}$ of the form
\begin{equation}
    \frac{\diff{\boldsymbol{\sf y}}}{\diff{t}} = g\pp{t, \hat\theta, \vec{\hat r},\boldsymbol{\sf y}}.
    \label{eqn: y vec evolution}
\end{equation}
Once instances of spatial derivatives have been replaced according to \eqref{eqn: derivative approximation}, the boundary conditions constitute conditions that $\boldsymbol{\sf y}$ must satisfy. Solving these equations for ${\sf y}_0, {\sf y}_1, {\sf y}_{N-1}, {\sf y}_N$ in terms of ${\sf y}_2, \dots,{\sf y}_{N-2}$ ensures that $\boldsymbol{\sf y}$ satisfies these conditions. 

Using the boundary conditions we can constrain ${\sf y}_0$, ${\sf y}_1$, ${\sf y}_{N-1}$ and ${\sf y}_N$. The first three boundary conditions, 
\begin{align}
   & {\sf y}(0, t) = 0,\quad 
  \pdiff{{\sf y}}{\sigma}(0, t) = 0,\quad\pdiffn{2}{{\sf y}}{\sigma}(L, t) = 0.  
\end{align}
are linear, which allows to express directly ${\sf y}_1, {\sf y}_{N-1}, {\sf y}_N$ in terms of ${\sf y}_0$, $ {\sf y}_2$, and ${\sf y}_{N-2}$. 
\begin{equation}
    B \pdiffn{3}{{\sf y}}{\sigma}(L, t) - \vec{F}  \cdot \boldsymbol{\hat \tau} \pdiff{{\sf y}}{\sigma}(L, t) = -\vec{F}  \cdot \boldsymbol{\hat \nu},
\end{equation}
can be discretised as
\begin{equation} \label{eqn:tip-bc-discretised}
     B \pp{ {2}/{L}}^3\left[D_N^3\boldsymbol{\sf y}\right]_0 - \pp{ {2}/{L}}\vec{F}  \cdot \boldsymbol{\hat \tau} \left[D_N\boldsymbol{\sf y}\right]_0 = -\vec{F}  \cdot \boldsymbol{\hat \nu};
\end{equation}
noting that 
$
       \linepdiffn{k}{{\sf y}}{X}=\pp{L/2}^k\linepdiffn{k}{{\sf y}}{\sigma}$ \eqref{eqn:change-variable-sigma-X}.
Here, $\vec{F}$ depends on the position of the tip, thus \eqref{eqn:tip-bc-discretised}  is a nonlinear equation for ${\sf y}_0$ for which no closed-form solution can be found in general. 
In principle, one can solve this nonlinear equation numerically for ${\sf y}_0$ using a root-finding procedure at each time step. Alternatively, we may simplify the problem by evaluating the tangent at the tip at the neighbouring point $X = X_2$ (instead of $X_1$), i.e.,
\begin{equation}
    \vec{t} = \displaystyle\frac{\linepdiff{\vec{r}}{\sigma}(L, t)}{ \left\|\linepdiff{\vec{r}}{\sigma}(L, t)\right\|} \approx \pp{{1 + \pp{\frac{2}{L}\frac{{\sf y}_2 - {\sf y}_3}{X_2 - X_3}}^2}} ^{-1/2}\pp{\boldsymbol{\hat \tau} + \frac{2}{L}\frac{{\sf y}_2 - {\sf y}_3}{X_2 - X_3}\boldsymbol{\hat \nu}},
    \label{eqn: approx t}
\end{equation}
so as to eliminate the explicit dependence upon ${\sf y}_0$.
The error produced by the approximation \eqref{eqn: approx t} for the tangent is $\mathcal{O}\pp{1/N^2}$ as $N \rightarrow \infty$. This converts the above equation to a linear one, which allows us to obtain a closed-form expression for ${\sf y}_0$ in terms of ${\sf y}_2, \dots,{\sf y}_{N-2}$.


Thus, \eqref{eqn:dimensional y equation} can be approximated by an equation of the form
\begin{equation}
    \frac{\diff{\boldsymbol{{\sf Y}}}}{\diff{t}} = G\pp{t, \hat\theta, \boldsymbol{\hat r},\boldsymbol{{\sf Y}}},
    \label{eqn: y vec evolution bcs}
\end{equation}
where $
    \boldsymbol{{\sf Y}} \eqdef \pp{{\sf y}_2, \dots, {\sf y}_{N-2}}$ denotes the vector of discrete variables after truncation of the boundary values ${\sf y_0}$, $ {\sf y_1} $, ${\sf  y_N}$ and ${\sf  y_{N-1}}$ (which are now fixed); and $ G(t, \hat\theta, \boldsymbol{\hat r},\boldsymbol{\sf Y}) \eqdef g(t, \hat\theta, \vec{\hat r},\boldsymbol{\sf y}(\boldsymbol{\sf Y}))$. 
Finally, \eqref{eqn:theta hat eqn} can be re-written as
\begin{equation}
    \fdiff{\hat{\theta}}{t} =\pp{ {2}/{L}}^2 V(t)\left[D_N^2\boldsymbol{\sf y}\right]_N,
    \label{eqn: theta hat approx}
\end{equation}
and \eqref{eqn:dimensional rHat equation} remains unchanged. 
Equations
\begin{align}
&\frac{\diff{\boldsymbol{{\sf Y}}}}{\diff{t}} = G\pp{t, \hat\theta, \boldsymbol{\hat r},\boldsymbol{{\sf Y}}},\\
    &\fdiff{\hat{\theta}}{t} =\pp{ {2}/{L}}^2 V(t)\left[D_N^2\boldsymbol{\sf y}\right]_N,\\
    &\fdiff{\vec{\hat r}}{t}  = V(t)\,  \boldsymbol{\hat \tau},
\end{align}
form a simple IVP which which can be integrated numerically (here we used the built-in function \textit{NDSolve} implemented in \textit{Mathematica 13.3}).

\bibliographystyle{elsarticle-harv}

\begin{thebibliography}{57}
\expandafter\ifx\csname natexlab\endcsname\relax\def\natexlab#1{#1}\fi
\providecommand{\url}[1]{\texttt{#1}}
\providecommand{\href}[2]{#2}
\providecommand{\path}[1]{#1}
\providecommand{\DOIprefix}{doi:}
\providecommand{\ArXivprefix}{arXiv:}
\providecommand{\URLprefix}{URL: }
\providecommand{\Pubmedprefix}{pmid:}
\providecommand{\doi}[1]{\href{http://dx.doi.org/#1}{\path{#1}}}
\providecommand{\Pubmed}[1]{\href{pmid:#1}{\path{#1}}}
\providecommand{\bibinfo}[2]{#2}
\ifx\xfnm\relax \def\xfnm[#1]{\unskip,\space#1}\fi
\bibitem[{Aeschlimann(2000)}]{Aeschlimann2000}
\bibinfo{author}{Aeschlimann, M.}, \bibinfo{year}{2000}.
\newblock \bibinfo{title}{{Biophysical model of axonal pathfinding}}.
\newblock Ph.D. thesis. University of Lausanne, Department of Theoretical Physics.
\bibitem[{Bangasser and Odde(2013)}]{bangasser2013master}
\bibinfo{author}{Bangasser, B.L.}, \bibinfo{author}{Odde, D.J.}, \bibinfo{year}{2013}.
\newblock \bibinfo{title}{Master equation-based analysis of a motor-clutch model for cell traction force}.
\newblock \bibinfo{journal}{Cellular and molecular bioengineering} \bibinfo{volume}{6}, \bibinfo{pages}{449--459}.
\newblock \URLprefix \url{https://doi.org/10.1007/s12195-013-0296-5}, \DOIprefix\doi{10.1007/s12195-013-0296-5}.
\bibitem[{Bangasser et~al.(2013)Bangasser, Rosenfeld and Odde}]{bangasser2013determinants}
\bibinfo{author}{Bangasser, B.L.}, \bibinfo{author}{Rosenfeld, S.S.}, \bibinfo{author}{Odde, D.J.}, \bibinfo{year}{2013}.
\newblock \bibinfo{title}{Determinants of maximal force transmission in a motor-clutch model of cell traction in a compliant microenvironment}.
\newblock \bibinfo{journal}{Biophysical journal} \bibinfo{volume}{105}, \bibinfo{pages}{581--592}.
\newblock \URLprefix \url{https://www.cell.com/biophysj/fulltext/S0006-3495(11)00376-6}, \DOIprefix\doi{10.1016/j.bpj.2013.06.027}.
\bibitem[{Bangasser et~al.(2017)Bangasser, Shamsan, Chan, Opoku, T{\"u}zel, Schlichtmann, Kasim, Fuller, McCullough, Rosenfeld et~al.}]{bangasser2017shifting}
\bibinfo{author}{Bangasser, B.L.}, \bibinfo{author}{Shamsan, G.A.}, \bibinfo{author}{Chan, C.E.}, \bibinfo{author}{Opoku, K.N.}, \bibinfo{author}{T{\"u}zel, E.}, \bibinfo{author}{Schlichtmann, B.W.}, \bibinfo{author}{Kasim, J.A.}, \bibinfo{author}{Fuller, B.J.}, \bibinfo{author}{McCullough, B.R.}, \bibinfo{author}{Rosenfeld, S.S.}, et~al., \bibinfo{year}{2017}.
\newblock \bibinfo{title}{Shifting the optimal stiffness for cell migration}.
\newblock \bibinfo{journal}{Nature Communications} \bibinfo{volume}{8}, \bibinfo{pages}{15313}.
\newblock \URLprefix \url{https://www.nature.com/articles/ncomms15313}, \DOIprefix\doi{10.1038/ncomms15313}.
\bibitem[{Bell(1978)}]{Bell1978}
\bibinfo{author}{Bell, G.I.}, \bibinfo{year}{1978}.
\newblock \bibinfo{title}{{Models for the specific adhesion of cells to cells}}.
\newblock \bibinfo{journal}{Science} \bibinfo{volume}{200}, \bibinfo{pages}{618--627}.
\newblock \URLprefix \url{https://www.science.org/doi/10.1126/science.347575}, \DOIprefix\doi{10.1126/SCIENCE.347575}.
\bibitem[{Bell et~al.(1984)Bell, Dembo and Bongrand}]{bell1984cell}
\bibinfo{author}{Bell, G.I.}, \bibinfo{author}{Dembo, M.}, \bibinfo{author}{Bongrand, P.}, \bibinfo{year}{1984}.
\newblock \bibinfo{title}{Cell adhesion. competition between nonspecific repulsion and specific bonding}.
\newblock \bibinfo{journal}{Biophysical journal} \bibinfo{volume}{45}, \bibinfo{pages}{1051--1064}.
\newblock \URLprefix \url{https://www.cell.com/biophysj/pdf/S0006-3495(84)84252-6.pdf}, \DOIprefix\doi{10.1016/S0006-3495(84)84252-6}.
\bibitem[{Bressloff(2020)}]{bressloff2020stochastic}
\bibinfo{author}{Bressloff, P.C.}, \bibinfo{year}{2020}.
\newblock \bibinfo{title}{Stochastic resetting and the mean-field dynamics of focal adhesions}.
\newblock \bibinfo{journal}{Physical Review E} \bibinfo{volume}{102}, \bibinfo{pages}{022134}.
\newblock \URLprefix \url{https://journals.aps.org/pre/abstract/10.1103/PhysRevE.102.022134}, \DOIprefix\doi{10.1103/PhysRevE.102.022134}.
\bibitem[{Chan and Odde(2008)}]{Chan2008}
\bibinfo{author}{Chan, C.E.}, \bibinfo{author}{Odde, D.J.}, \bibinfo{year}{2008}.
\newblock \bibinfo{title}{{Traction dynamics of filopodia on compliant substrates}}.
\newblock \bibinfo{journal}{Science} \bibinfo{volume}{322}, \bibinfo{pages}{1687--1691}.
\newblock \URLprefix \url{https://www.science.org/doi/10.1126/science.1163595}, \DOIprefix\doi{10.1126/science.1163595}.
\bibitem[{Chaudhuri et~al.(2011)Chaudhuri, Borowski and Zapotocky}]{chaudhuri2011model}
\bibinfo{author}{Chaudhuri, D.}, \bibinfo{author}{Borowski, P.}, \bibinfo{author}{Zapotocky, M.}, \bibinfo{year}{2011}.
\newblock \bibinfo{title}{Model of fasciculation and sorting in mixed populations of axons}.
\newblock \bibinfo{journal}{Physical Review E—Statistical, Nonlinear, and Soft Matter Physics} \bibinfo{volume}{84}, \bibinfo{pages}{021908}.
\newblock \URLprefix \url{https://journals.aps.org/pre/abstract/10.1103/PhysRevE.84.021908}, \DOIprefix\doi{10.1103/PhysRevE.84.021908}.
\bibitem[{De~Gennes(2007)}]{de2007collective}
\bibinfo{author}{De~Gennes, P.G.}, \bibinfo{year}{2007}.
\newblock \bibinfo{title}{Collective neuronal growth and self organization of axons}.
\newblock \bibinfo{journal}{Proceedings of the National Academy of Sciences} \bibinfo{volume}{104}, \bibinfo{pages}{4904--4906}.
\newblock \URLprefix \url{https://www.pnas.org/doi/full/10.1073/pnas.0609871104}, \DOIprefix\doi{10.1073/pnas.0609871104}.
\bibitem[{Dennerll et~al.(1989)Dennerll, Lamoureux, Buxbaum and Heidemann}]{Dennerll1989}
\bibinfo{author}{Dennerll, T.J.}, \bibinfo{author}{Lamoureux, P.}, \bibinfo{author}{Buxbaum, R.E.}, \bibinfo{author}{Heidemann, S.R.}, \bibinfo{year}{1989}.
\newblock \bibinfo{title}{{The cytomechanics of axonal elongation and retraction}}.
\newblock \bibinfo{journal}{Journal of Cell Biology} \bibinfo{volume}{109}, \bibinfo{pages}{3073--3083}.
\newblock \URLprefix \url{https://rupress.org/jcb/article-abstract/109/6/3073/28999}, \DOIprefix\doi{10.1083/jcb.109.6.3073}.
\bibitem[{Elosegui-Artola et~al.(2018)Elosegui-Artola, Trepat and Roca-Cusachs}]{elosegui2018control}
\bibinfo{author}{Elosegui-Artola, A.}, \bibinfo{author}{Trepat, X.}, \bibinfo{author}{Roca-Cusachs, P.}, \bibinfo{year}{2018}.
\newblock \bibinfo{title}{Control of mechanotransduction by molecular clutch dynamics}.
\newblock \bibinfo{journal}{Trends in Cell Biology} \bibinfo{volume}{28}, \bibinfo{pages}{356--367}.
\newblock \URLprefix \url{https://www.cell.com/trends/cell-biology/fulltext/S0962-8924(18)30017-5}, \DOIprefix\doi{10.1016/j.tcb.2018.01.008}.
\bibitem[{Espina et~al.(2022)Espina, Marchant and Barriga}]{espina2022durotaxis}
\bibinfo{author}{Espina, J.A.}, \bibinfo{author}{Marchant, C.L.}, \bibinfo{author}{Barriga, E.H.}, \bibinfo{year}{2022}.
\newblock \bibinfo{title}{Durotaxis: the mechanical control of directed cell migration}.
\newblock \bibinfo{journal}{The FEBS journal} \bibinfo{volume}{289}, \bibinfo{pages}{2736--2754}.
\newblock \URLprefix \url{https://febs.onlinelibrary.wiley.com/doi/10.1111/febs.15862}, \DOIprefix\doi{10.1111/febs.15862}.
\bibitem[{Franze(2020)}]{Franze2020}
\bibinfo{author}{Franze, K.}, \bibinfo{year}{2020}.
\newblock \bibinfo{title}{{Integrating Chemistry and Mechanics: The Forces Driving Axon Growth}}.
\newblock \bibinfo{journal}{Annual Review of Cell and Developmental Biology} \bibinfo{volume}{36}, \bibinfo{pages}{61--83}.
\newblock \URLprefix \url{https://www.annualreviews.org/content/journals/10.1146/annurev-cellbio-100818-125157}, \DOIprefix\doi{10.1146/annurev-cellbio-100818-125157}.
\bibitem[{Franze and Guck(2010)}]{Franze2010}
\bibinfo{author}{Franze, K.}, \bibinfo{author}{Guck, J.}, \bibinfo{year}{2010}.
\newblock \bibinfo{title}{{The biophysics of neuronal growth}}.
\newblock \bibinfo{journal}{Reports on Progress in Physics} \bibinfo{volume}{73}.
\newblock \URLprefix \url{https://doi.org/10.1088/0034-4885/73/9/094601}, \DOIprefix\doi{10.1088/0034-4885/73/9/094601}.
\bibitem[{Goriely(2017)}]{Goriely2017}
\bibinfo{author}{Goriely, A.}, \bibinfo{year}{2017}.
\newblock \bibinfo{title}{{The mathematics and mechanics of biological growth}}. volume~\bibinfo{volume}{45} of \textit{\bibinfo{series}{Interdisciplinary applied mathematics}}.
\newblock \bibinfo{edition}{1st} ed., \bibinfo{publisher}{Springer-Verlag, New York}.
\newblock \URLprefix \url{https://link.springer.com/book/10.1007/978-0-387-87710-5}, \DOIprefix\doi{10.1007/978-0-387-87710-5}.
\bibitem[{Goriely et~al.(2015)Goriely, Budday and Kuhl}]{Goriely2015}
\bibinfo{author}{Goriely, A.}, \bibinfo{author}{Budday, S.}, \bibinfo{author}{Kuhl, E.}, \bibinfo{year}{2015}.
\newblock \bibinfo{title}{{Neuromechanics: From neurons to brain}}, in: \bibinfo{booktitle}{Advances in Applied Mechanics}. \bibinfo{publisher}{Academic Press Inc.}. volume~\bibinfo{volume}{48}, pp. \bibinfo{pages}{79--139}.
\newblock \URLprefix \url{https://doi.org/10.1016/bs.aams.2015.10.002}, \DOIprefix\doi{10.1016/bs.aams.2015.10.002}.
\bibitem[{Goriely et~al.(2023)Goriely, Moulton and Mihai}]{goriely2023rod}
\bibinfo{author}{Goriely, A.}, \bibinfo{author}{Moulton, D.E.}, \bibinfo{author}{Mihai, L.A.}, \bibinfo{year}{2023}.
\newblock \bibinfo{title}{A rod theory for liquid crystalline elastomers}.
\newblock \bibinfo{journal}{Journal of Elasticity} \bibinfo{volume}{153}, \bibinfo{pages}{509--532}.
\newblock \URLprefix \url{https://link.springer.com/article/10.1007/s10659-021-09875-z}, \DOIprefix\doi{10.1007/s10659-021-09875}.
\bibitem[{Hentschel and {Van Ooyen}(1999)}]{Hentschel1999}
\bibinfo{author}{Hentschel, H.G.}, \bibinfo{author}{{Van Ooyen}, A.}, \bibinfo{year}{1999}.
\newblock \bibinfo{title}{{Models of axon guidance and bundling during development}}.
\newblock \bibinfo{journal}{Proceedings of the Royal Society B: Biological Sciences} \bibinfo{volume}{266}, \bibinfo{pages}{2231--2238}.
\newblock \URLprefix \url{https://doi.org/10.1098/rspb.1999.0913}, \DOIprefix\doi{10.1098/rspb.1999.0913}.
\bibitem[{Holland et~al.(2015)Holland, Miller and Kuhl}]{Holland2015}
\bibinfo{author}{Holland, M.A.}, \bibinfo{author}{Miller, K.E.}, \bibinfo{author}{Kuhl, E.}, \bibinfo{year}{2015}.
\newblock \bibinfo{title}{{Emerging Brain Morphologies from Axonal Elongation}}.
\newblock \bibinfo{journal}{Annals of Biomedical Engineering} \bibinfo{volume}{43}, \bibinfo{pages}{1640--1653}.
\newblock \URLprefix \url{https://doi.org/10.1007/s10439-015-1312-9}, \DOIprefix\doi{10.1007/s10439-015-1312-9}.
\bibitem[{Hwu(2010)}]{Hwu2010}
\bibinfo{author}{Hwu, C.}, \bibinfo{year}{2010}.
\newblock \bibinfo{title}{Infinite Space, Half-Space, and Bimaterials}. \bibinfo{publisher}{Springer US}, \bibinfo{address}{Boston, MA}. chapter~\bibinfo{chapter}{4}.
\newblock pp. \bibinfo{pages}{87--113}.
\newblock \URLprefix \url{https://doi.org/10.1007/978-1-4419-5915-7_4}, \DOIprefix\doi{10.1007/978-1-4419-5915-7_4}.
\bibitem[{Isomursu et~al.(2022)Isomursu, Park, Hou, Cheng, Mathieu, Shamsan, Fuller, Kasim, Mahmoodi, Lu, Genin, Xu, Lin, Distefano, Ivaska and Odde}]{isomursu2022directed}
\bibinfo{author}{Isomursu, A.}, \bibinfo{author}{Park, K.Y.}, \bibinfo{author}{Hou, J.}, \bibinfo{author}{Cheng, B.}, \bibinfo{author}{Mathieu, M.}, \bibinfo{author}{Shamsan, G.A.}, \bibinfo{author}{Fuller, B.}, \bibinfo{author}{Kasim, J.}, \bibinfo{author}{Mahmoodi, M.M.}, \bibinfo{author}{Lu, T.J.}, \bibinfo{author}{Genin, G.M.}, \bibinfo{author}{Xu, F.}, \bibinfo{author}{Lin, M.}, \bibinfo{author}{Distefano, M.D.}, \bibinfo{author}{Ivaska, J.}, \bibinfo{author}{Odde, D.J.}, \bibinfo{year}{2022}.
\newblock \bibinfo{title}{Directed cell migration towards softer environments}.
\newblock \bibinfo{journal}{Nature Materials} \bibinfo{volume}{21}, \bibinfo{pages}{1081–1090}.
\newblock \URLprefix \url{https://www.nature.com/articles/s41563-022-01294-2}, \DOIprefix\doi{10.1038/s41563-022-01294-2}.
\bibitem[{Koser et~al.(2016)Koser, Thompson, Foster, Dwivedy, Pillai, Sheridan, Svoboda, Viana, da~Fontoura~Costa, Guck, Holt and Franze}]{koser2016mechanosensing}
\bibinfo{author}{Koser, D.E.}, \bibinfo{author}{Thompson, A.J.}, \bibinfo{author}{Foster, S.K.}, \bibinfo{author}{Dwivedy, A.}, \bibinfo{author}{Pillai, E.K.}, \bibinfo{author}{Sheridan, G.K.}, \bibinfo{author}{Svoboda, H.}, \bibinfo{author}{Viana, M.}, \bibinfo{author}{da~Fontoura~Costa, L.}, \bibinfo{author}{Guck, J.}, \bibinfo{author}{Holt, C.E.}, \bibinfo{author}{Franze, K.}, \bibinfo{year}{2016}.
\newblock \bibinfo{title}{Mechanosensing is critical for axon growth in the developing brain}.
\newblock \bibinfo{journal}{Nature neuroscience} \bibinfo{volume}{19}, \bibinfo{pages}{1592}.
\newblock \URLprefix \url{https://doi.org/10.1038/nn.4394}, \DOIprefix\doi{https://doi.org/10.1038/nn.4394}.
\bibitem[{Kov{\'a}cs et~al.(2004)Kov{\'a}cs, T{\'o}th, Het{\'e}nyi, M{\'a}ln{\'a}si-Csizmadia and Sellers}]{kovacs2004mechanism}
\bibinfo{author}{Kov{\'a}cs, M.}, \bibinfo{author}{T{\'o}th, J.}, \bibinfo{author}{Het{\'e}nyi, C.}, \bibinfo{author}{M{\'a}ln{\'a}si-Csizmadia, A.}, \bibinfo{author}{Sellers, J.R.}, \bibinfo{year}{2004}.
\newblock \bibinfo{title}{{Mechanism of blebbistatin inhibition of myosin II}}.
\newblock \bibinfo{journal}{Journal of Biological Chemistry} \bibinfo{volume}{279}, \bibinfo{pages}{35557--35563}.
\newblock \URLprefix \url{https://www.sciencedirect.com/science/article/pii/S0021925820731376}, \DOIprefix\doi{10.1074/jbc.M405319200}.
\bibitem[{Lamoureux et~al.(1989)Lamoureux, Buxbaum and Heidemann}]{Lamoureux1989}
\bibinfo{author}{Lamoureux, P.}, \bibinfo{author}{Buxbaum, R.E.}, \bibinfo{author}{Heidemann, S.R.}, \bibinfo{year}{1989}.
\newblock \bibinfo{title}{{Direct evidence that growth cones pull}}.
\newblock \bibinfo{journal}{Nature} \bibinfo{volume}{340}, \bibinfo{pages}{159--162}.
\newblock \URLprefix \url{http://www.nature.com/articles/340159a0}, \DOIprefix\doi{10.1038/340159a0}.
\bibitem[{Lin et~al.(1996)Lin, Espreafico, Mooseker and Forscher}]{lin1996myosin}
\bibinfo{author}{Lin, C.H.}, \bibinfo{author}{Espreafico, E.M.}, \bibinfo{author}{Mooseker, M.S.}, \bibinfo{author}{Forscher, P.}, \bibinfo{year}{1996}.
\newblock \bibinfo{title}{Myosin drives retrograde f-actin flow in neuronal growth cones}.
\newblock \bibinfo{journal}{Neuron} \bibinfo{volume}{16}, \bibinfo{pages}{769--782}.
\newblock \URLprefix \url{https://doi.org/10.1016/s0896-6273(00)80097-5}, \DOIprefix\doi{10.1016/s0896-6273(00)80097-5}.
\bibitem[{Maskery and Shinbrot(2005)}]{maskery2005deterministic}
\bibinfo{author}{Maskery, S.}, \bibinfo{author}{Shinbrot, T.}, \bibinfo{year}{2005}.
\newblock \bibinfo{title}{Deterministic and stochastic elements of axonal guidance}.
\newblock \bibinfo{journal}{Annual Review of Biomedical Engineering} \bibinfo{volume}{7}, \bibinfo{pages}{187--221}.
\newblock \URLprefix \url{https://www.annualreviews.org/content/journals/10.1146/annurev.bioeng.7.060804.100446}, \DOIprefix\doi{10.1146/annurev.bioeng.7.060804.100446}.
\bibitem[{McCormick and Gupton(2020)}]{mccormick2020mechanistic}
\bibinfo{author}{McCormick, L.E.}, \bibinfo{author}{Gupton, S.L.}, \bibinfo{year}{2020}.
\newblock \bibinfo{title}{Mechanistic advances in axon pathfinding}.
\newblock \bibinfo{journal}{Current Opinion in Cell Biology} \bibinfo{volume}{63}, \bibinfo{pages}{11--19}.
\newblock \URLprefix \url{https://www.sciencedirect.com/science/article/pii/S0955067419301139}, \DOIprefix\doi{10.1016/j.ceb.2019.12.003}.
\bibitem[{Miller and Suter(2018)}]{miller2018integrated}
\bibinfo{author}{Miller, K.E.}, \bibinfo{author}{Suter, D.M.}, \bibinfo{year}{2018}.
\newblock \bibinfo{title}{An integrated cytoskeletal model of neurite outgrowth}.
\newblock \bibinfo{journal}{Frontiers in cellular neuroscience} \bibinfo{volume}{12}, \bibinfo{pages}{447}.
\newblock \URLprefix \url{https://doi.org/10.3389/fncel.2018.00447}, \DOIprefix\doi{10.3389/fncel.2018.00447}.
\bibitem[{Mortimer et~al.(2008)Mortimer, Fothergill, Pujic, Richards and Goodhill}]{Mortimer2008}
\bibinfo{author}{Mortimer, D.}, \bibinfo{author}{Fothergill, T.}, \bibinfo{author}{Pujic, Z.}, \bibinfo{author}{Richards, L.J.}, \bibinfo{author}{Goodhill, G.J.}, \bibinfo{year}{2008}.
\newblock \bibinfo{title}{Growth cone chemotaxis}.
\newblock \bibinfo{journal}{Trends in Neurosciences} \bibinfo{volume}{31}, \bibinfo{pages}{90--98}.
\newblock \URLprefix \url{https://doi.org/10.1016/j.tins.2007.11.008}, \DOIprefix\doi{10.1016/j.tins.2007.11.008}.
\bibitem[{Moulton et~al.(2020)Moulton, Oliveri and Goriely}]{moulton2020multiscale}
\bibinfo{author}{Moulton, D.E.}, \bibinfo{author}{Oliveri, H.}, \bibinfo{author}{Goriely, A.}, \bibinfo{year}{2020}.
\newblock \bibinfo{title}{Multiscale integration of environmental stimuli in plant tropism produces complex behaviors}.
\newblock \bibinfo{journal}{Proceedings of the National Academy of Sciences} \bibinfo{volume}{117}, \bibinfo{pages}{32226--32237}.
\newblock \URLprefix \url{https://doi.org/10.1073/pnas.2016025117}, \DOIprefix\doi{10.1073/pnas.2016025117}.
\bibitem[{Nguyen et~al.(2013)Nguyen, Hogue, Cung, Purohit and McAlpine}]{nguyen2013tension}
\bibinfo{author}{Nguyen, T.D.}, \bibinfo{author}{Hogue, I.B.}, \bibinfo{author}{Cung, K.}, \bibinfo{author}{Purohit, P.K.}, \bibinfo{author}{McAlpine, M.C.}, \bibinfo{year}{2013}.
\newblock \bibinfo{title}{Tension-induced neurite growth in microfluidic channels}.
\newblock \bibinfo{journal}{Lab on a Chip} \bibinfo{volume}{13}, \bibinfo{pages}{3735--3740}.
\newblock \URLprefix \url{https://pubs.rsc.org/en/content/articlelanding/2013/lc/c3lc50681a}, \DOIprefix\doi{10.1039/C3LC50681A}.
\bibitem[{Nicolas et~al.(2008)Nicolas, Besser and Safran}]{nicolas2008dynamics}
\bibinfo{author}{Nicolas, A.}, \bibinfo{author}{Besser, A.}, \bibinfo{author}{Safran, S.A.}, \bibinfo{year}{2008}.
\newblock \bibinfo{title}{Dynamics of cellular focal adhesions on deformable substrates: consequences for cell force microscopy}.
\newblock \bibinfo{journal}{Biophysical journal} \bibinfo{volume}{95}, \bibinfo{pages}{527--539}.
\newblock \URLprefix \url{https://doi.org/10.1529/biophysj.107.127399}, \DOIprefix\doi{10.1529/biophysj.107.127399}.
\bibitem[{Oliveri et~al.(2022)Oliveri, De~Rooij, Kuhl and Goriely}]{Oliveri2022b}
\bibinfo{author}{Oliveri, H.}, \bibinfo{author}{De~Rooij, R.}, \bibinfo{author}{Kuhl, E.}, \bibinfo{author}{Goriely, A.}, \bibinfo{year}{2022}.
\newblock \bibinfo{title}{Rheology of growing axons}.
\newblock \bibinfo{journal}{Physical Review Research} \bibinfo{volume}{4}, \bibinfo{pages}{033125}.
\newblock \URLprefix \url{https://link.aps.org/doi/10.1103/PhysRevResearch.4.033125}, \DOIprefix\doi{10.1103/PhysRevResearch.4.033125}.
\bibitem[{Oliveri et~al.(2021)Oliveri, Franze and Goriely}]{Oliveri2021}
\bibinfo{author}{Oliveri, H.}, \bibinfo{author}{Franze, K.}, \bibinfo{author}{Goriely, A.}, \bibinfo{year}{2021}.
\newblock \bibinfo{title}{{Theory for Durotactic Axon Guidance}}.
\newblock \bibinfo{journal}{Physical Review Letters} \bibinfo{volume}{126}, \bibinfo{pages}{118101}.
\newblock \URLprefix \url{https://doi.org/10.1103/PhysRevLett.126.118101}, \DOIprefix\doi{10.1103/PhysRevLett.126.118101}.
\bibitem[{Oliveri and Goriely(2022)}]{Oliveri2022a}
\bibinfo{author}{Oliveri, H.}, \bibinfo{author}{Goriely, A.}, \bibinfo{year}{2022}.
\newblock \bibinfo{title}{{Mathematical models of neuronal growth}}.
\newblock \bibinfo{journal}{Biomechanics and Modeling in Mechanobiology} \bibinfo{volume}{21}, \bibinfo{pages}{89--118}.
\newblock \URLprefix \url{https://link.springer.com/article/10.1007/s10237-021-01539-0}, \DOIprefix\doi{10.1007/S10237-021-01539-0}.
\bibitem[{O'Toole et~al.(2008)O'Toole, Lamoureux and Miller}]{OToole2008}
\bibinfo{author}{O'Toole, M.}, \bibinfo{author}{Lamoureux, P.}, \bibinfo{author}{Miller, K.E.}, \bibinfo{year}{2008}.
\newblock \bibinfo{title}{{A physical model of axonal elongation: Force, viscosity, and adhesions govern the mode of outgrowth}}.
\newblock \bibinfo{journal}{Biophysical Journal} \bibinfo{volume}{94}, \bibinfo{pages}{2610--2620}.
\newblock \URLprefix \url{https://www.cell.com/biophysj/fulltext/S0006-3495(08)70514-9}, \DOIprefix\doi{10.1529/biophysj.107.117424}.
\bibitem[{O'Toole and Miller(2011)}]{OToole2011}
\bibinfo{author}{O'Toole, M.}, \bibinfo{author}{Miller, K.E.}, \bibinfo{year}{2011}.
\newblock \bibinfo{title}{{The role of stretching in slow axonal transport}}.
\newblock \bibinfo{journal}{Biophysical Journal} \bibinfo{volume}{100}, \bibinfo{pages}{351--360}.
\newblock \URLprefix \url{https://www.cell.com/biophysj/fulltext/S0006-3495(10)05210-0}, \DOIprefix\doi{10.1016/j.bpj.2010.12.3695}.
\bibitem[{Press et~al.(2007)Press, Teukolsky, Vetterling and Flannery}]{Press2007}
\bibinfo{author}{Press, W.H.}, \bibinfo{author}{Teukolsky, S.A.}, \bibinfo{author}{Vetterling, W.T.}, \bibinfo{author}{Flannery, B.P.}, \bibinfo{year}{2007}.
\newblock \bibinfo{title}{Numerical recipes: The art of scientific computing}.
\newblock \bibinfo{edition}{3rd} ed., \bibinfo{publisher}{Cambridge University Press}, \bibinfo{address}{New York}.
\newblock \URLprefix \url{https://www.cambridge.org/de/universitypress/subjects/mathematics/numerical-recipes/numerical-recipes-art-scientific-computing-3rd-edition}.
\bibitem[{Purohit and Smith(2016)}]{purohit2016model}
\bibinfo{author}{Purohit, P.K.}, \bibinfo{author}{Smith, D.H.}, \bibinfo{year}{2016}.
\newblock \bibinfo{title}{A model for stretch growth of neurons}.
\newblock \bibinfo{journal}{Journal of biomechanics} \bibinfo{volume}{49}, \bibinfo{pages}{3934--3942}.
\newblock \URLprefix \url{https://www.sciencedirect.com/science/article/pii/S0021929016312246}, \DOIprefix\doi{10.1016/j.jbiomech.2016.11.045}.
\bibitem[{Raffa(2023)}]{RAFFA2022}
\bibinfo{author}{Raffa, V.}, \bibinfo{year}{2023}.
\newblock \bibinfo{title}{Force: A messenger of axon outgrowth}.
\newblock \bibinfo{journal}{Seminars in Cell \& Developmental Biology} \bibinfo{volume}{140}, \bibinfo{pages}{3--12}.
\newblock \URLprefix \url{https://doi.org/10.1016/j.semcdb.2022.07.004}, \DOIprefix\doi{10.1016/j.semcdb.2022.07.004}. \bibinfo{note}{special issue: Driving forces behind the wiring of neuronal circuits}.
\bibitem[{Recho et~al.(2016)Recho, J\'erusalem and Goriely}]{Recho2016}
\bibinfo{author}{Recho, P.}, \bibinfo{author}{J\'erusalem, A.}, \bibinfo{author}{Goriely, A.}, \bibinfo{year}{2016}.
\newblock \bibinfo{title}{Growth, collapse, and stalling in a mechanical model for neurite motility}.
\newblock \bibinfo{journal}{Physical Review E} \bibinfo{volume}{93}.
\newblock \URLprefix \url{https://journals.aps.org/pre/abstract/10.1103/PhysRevE.93.032410}, \DOIprefix\doi{10.1103/PhysRevE.93.032410}.
\bibitem[{Sabass and Schwarz(2010)}]{sabass2010modeling}
\bibinfo{author}{Sabass, B.}, \bibinfo{author}{Schwarz, U.S.}, \bibinfo{year}{2010}.
\newblock \bibinfo{title}{Modeling cytoskeletal flow over adhesion sites: competition between stochastic bond dynamics and intracellular relaxation}.
\newblock \bibinfo{journal}{Journal of Physics: Condensed Matter} \bibinfo{volume}{22}, \bibinfo{pages}{194112}.
\newblock \URLprefix \url{https://doi.org/10.1088/0953-8984/22/19/194112}, \DOIprefix\doi{10.1088/0953-8984/22/19/194112}.
\bibitem[{S{\'a}ez and Venturini(2023)}]{saez2023positive}
\bibinfo{author}{S{\'a}ez, P.}, \bibinfo{author}{Venturini, C.}, \bibinfo{year}{2023}.
\newblock \bibinfo{title}{Positive{,} negative and controlled durotaxis}.
\newblock \bibinfo{journal}{Soft Matter} \bibinfo{volume}{19}, \bibinfo{pages}{2993--3001}.
\newblock \URLprefix \url{https://doi.org/10.1039/D2SM01326F}, \DOIprefix\doi{10.1039/D2SM01326F}.
\bibitem[{Santos et~al.(2020)Santos, Schaffran, Broguière, Meyn, Zenobi-Wong and Bradke}]{SANTOS2020107907}
\bibinfo{author}{Santos, T.E.}, \bibinfo{author}{Schaffran, B.}, \bibinfo{author}{Broguière, N.}, \bibinfo{author}{Meyn, L.}, \bibinfo{author}{Zenobi-Wong, M.}, \bibinfo{author}{Bradke, F.}, \bibinfo{year}{2020}.
\newblock \bibinfo{title}{Axon growth of cns neurons in three dimensions is amoeboid and independent of adhesions}.
\newblock \bibinfo{journal}{Cell Reports} \bibinfo{volume}{32}, \bibinfo{pages}{107907}.
\newblock \URLprefix \url{https://www.sciencedirect.com/science/article/pii/S2211124720308883}, \DOIprefix\doi{https://doi.org/10.1016/j.celrep.2020.107907}.
\bibitem[{Schaeffer et~al.(2022)Schaeffer, Weber, Thompson, Keynes and Franze}]{Schaeffer2022}
\bibinfo{author}{Schaeffer, J.}, \bibinfo{author}{Weber, I.P.}, \bibinfo{author}{Thompson, A.J.}, \bibinfo{author}{Keynes, R.J.}, \bibinfo{author}{Franze, K.}, \bibinfo{year}{2022}.
\newblock \bibinfo{title}{Axons in the chick embryo follow soft pathways through developing somite segments}.
\newblock \bibinfo{journal}{Frontiers in Cell and Developmental Biology} \bibinfo{volume}{10}.
\newblock \URLprefix \url{https://www.frontiersin.org/articles/10.3389/fcell.2022.917589}, \DOIprefix\doi{10.3389/fcell.2022.917589}.
\bibitem[{Sens(2013)}]{sens2013rigidity}
\bibinfo{author}{Sens, P.}, \bibinfo{year}{2013}.
\newblock \bibinfo{title}{Rigidity sensing by stochastic sliding friction}.
\newblock \bibinfo{journal}{Europhysics Letters} \bibinfo{volume}{104}, \bibinfo{pages}{38003}.
\newblock \URLprefix \url{https://iopscience.iop.org/article/10.1209/0295-5075/104/38003}, \DOIprefix\doi{10.1209/0295-5075/104/38003}.
\bibitem[{Shellard and Mayor(2021)}]{SHELLARD2021227}
\bibinfo{author}{Shellard, A.}, \bibinfo{author}{Mayor, R.}, \bibinfo{year}{2021}.
\newblock \bibinfo{title}{Durotaxis: The hard path from in vitro to in vivo}.
\newblock \bibinfo{journal}{Developmental Cell} \bibinfo{volume}{56}, \bibinfo{pages}{227--239}.
\newblock \URLprefix \url{https://www.sciencedirect.com/science/article/pii/S153458072030928X}, \DOIprefix\doi{10.1016/j.devcel.2020.11.019}.
\bibitem[{Silk and Erickson(1979)}]{SILK1979481}
\bibinfo{author}{Silk, W.K.}, \bibinfo{author}{Erickson, R.O.}, \bibinfo{year}{1979}.
\newblock \bibinfo{title}{Kinematics of plant growth}.
\newblock \bibinfo{journal}{Journal of Theoretical Biology} \bibinfo{volume}{76}, \bibinfo{pages}{481--501}.
\newblock \URLprefix \url{https://www.sciencedirect.com/science/article/pii/0022519379900146}, \DOIprefix\doi{https://doi.org/10.1016/0022-5193(79)90014-6}.
\bibitem[{Smeal et~al.(2005)Smeal, Rabbitt, Biran and Tresco}]{smeal2005substrate}
\bibinfo{author}{Smeal, R.M.}, \bibinfo{author}{Rabbitt, R.}, \bibinfo{author}{Biran, R.}, \bibinfo{author}{Tresco, P.A.}, \bibinfo{year}{2005}.
\newblock \bibinfo{title}{Substrate curvature influences the direction of nerve outgrowth}.
\newblock \bibinfo{journal}{Annals of biomedical engineering} \bibinfo{volume}{33}, \bibinfo{pages}{376--382}.
\newblock \URLprefix \url{https://link.springer.com/article/10.1007/s10439-005-1740-z}, \DOIprefix\doi{10.1007/s10439-005-1740-z}.
\bibitem[{{\v{S}}m{\'{i}}t et~al.(2017){\v{S}}m{\'{i}}t, Fouquet, Pincet, Zapotocky and Trembleau}]{Smit2017}
\bibinfo{author}{{\v{S}}m{\'{i}}t, D.}, \bibinfo{author}{Fouquet, C.}, \bibinfo{author}{Pincet, F.}, \bibinfo{author}{Zapotocky, M.}, \bibinfo{author}{Trembleau, A.}, \bibinfo{year}{2017}.
\newblock \bibinfo{title}{{Axon tension regulates fasciculation/defasciculation through the control of axon shaft zippering}}.
\newblock \bibinfo{journal}{eLife} \bibinfo{volume}{6}.
\newblock \URLprefix \url{https://elifesciences.org/articles/19907}, \DOIprefix\doi{10.7554/eLife.19907}.
\bibitem[{Srinivasan and Walcott(2009)}]{srinivasan2009binding}
\bibinfo{author}{Srinivasan, M.}, \bibinfo{author}{Walcott, S.}, \bibinfo{year}{2009}.
\newblock \bibinfo{title}{Binding site models of friction due to the formation and rupture of bonds: state-function formalism, force-velocity relations, response to slip velocity transients, and slip stability}.
\newblock \bibinfo{journal}{Physical Review E} \bibinfo{volume}{80}, \bibinfo{pages}{046124}.
\newblock \URLprefix \url{https://journals.aps.org/pre/abstract/10.1103/PhysRevE.80.046124}, \DOIprefix\doi{10.1103/PhysRevE.80.046124}.
\bibitem[{Suter and Miller(2011)}]{Suter2011}
\bibinfo{author}{Suter, D.M.}, \bibinfo{author}{Miller, K.E.}, \bibinfo{year}{2011}.
\newblock \bibinfo{title}{{The emerging role of forces in axonal elongation}}.
\newblock \bibinfo{journal}{Progress in Neurobiology} \bibinfo{volume}{94}, \bibinfo{pages}{91--101}.
\newblock \URLprefix \url{https://www.sciencedirect.com/science/article/pii/S0301008211000530}, \DOIprefix\doi{10.1016/j.pneurobio.2011.04.002}.
\bibitem[{Thompson et~al.(2019)Thompson, Pillai, Dimov, Foster, Holt and Franze}]{Thompson2019}
\bibinfo{author}{Thompson, A.J.}, \bibinfo{author}{Pillai, E.K.}, \bibinfo{author}{Dimov, I.B.}, \bibinfo{author}{Foster, S.K.}, \bibinfo{author}{Holt, C.E.}, \bibinfo{author}{Franze, K.}, \bibinfo{year}{2019}.
\newblock \bibinfo{title}{Rapid changes in tissue mechanics regulate cell behaviour in the developing embryonic brain}.
\newblock \bibinfo{journal}{eLife} \bibinfo{volume}{8}, \bibinfo{pages}{e39356}.
\newblock \URLprefix \url{https://elifesciences.org/articles/39356}, \DOIprefix\doi{10.7554/eLife.39356}.
\bibitem[{Trefethen(2000)}]{trefethen2000spectral}
\bibinfo{author}{Trefethen, L.N.}, \bibinfo{year}{2000}.
\newblock \bibinfo{title}{Spectral methods in MATLAB}.
\newblock \bibinfo{publisher}{SIAM}.
\newblock \URLprefix \url{https://epubs.siam.org/doi/book/10.1137/1.9780898719598}, \DOIprefix\doi{10.1137/1.9780898719598}.
\bibitem[{Turney and Bridgman(2005)}]{turney2005laminin}
\bibinfo{author}{Turney, S.G.}, \bibinfo{author}{Bridgman, P.C.}, \bibinfo{year}{2005}.
\newblock \bibinfo{title}{{Laminin stimulates and guides axonal outgrowth via growth cone myosin II activity}}.
\newblock \bibinfo{journal}{Nature neuroscience} \bibinfo{volume}{8}, \bibinfo{pages}{717--719}.
\newblock \URLprefix \url{https://www.nature.com/articles/nn1466}, \DOIprefix\doi{10.1038/nn1466}.
\bibitem[{Wolpert(2008)}]{wolpert2017french}
\bibinfo{author}{Wolpert, L.}, \bibinfo{year}{2008}.
\newblock \bibinfo{title}{{The French Flag Problems A Contribution to the Discussion on Pattern Development and Regulation}}, in: \bibinfo{editor}{Waddington, C.H.} (Ed.), \bibinfo{booktitle}{The Origin of Life}. \bibinfo{publisher}{Routledge}. volume~\bibinfo{volume}{1}, pp. \bibinfo{pages}{125--133}.
\newblock \URLprefix \url{https://www.taylorfrancis.com/chapters/edit/10.4324/9781315133638-12/}.

\end{thebibliography}

\end{document}